# Polarization patterning in ferroelectric nematic liquids

N. Sebastián[1], M. Lovšin[1,2], B. Berteloot[3], N. Osterman[1,2], A. Petelin[1,2], R. J. Mandle[4,5], S. Aya[6,7], M. Huang[6,7], I. Drevenšek-Olenik[1,2], K. Neyts[2], A. Mertelj[1]

1 Jožef Stefan Institute, Ljubljana, Slovenia
2 University of Ljubljana, Faculty of Mathematics and Physics, Ljubljana, Slovenia
3 Liquid Crystals and Photonics Group, ELIS Department, Ghent University, Ghent, Belgium
4 School of Physics and Astronomy, University of Leeds, Leeds, UK
5 School of Chemistry, University of Leeds, Leeds, UK
6 South China Advanced Institute for Soft Matter Science and Technology (AISMST), School of Emergent Soft Matter, South China University of Technology, Guangzhou, China
7 Guangdong Provincial Key Laboratory of Functional and Intelligent Hybrid Materials and Devices, South China University of Technology, Guangzhou, China

**The recently discovered ferroelectric nematic liquids incorporate to the functional combination of fluidity, processability and anisotropic optical properties of nematic liquids, an astonishing range of physical properties derived from the phase polarity. Among them, the remarkably large values of second order optical susceptibility encourage to exploit these new materials for non-linear photonic applications. Here we show that photopatterning of the alignment layer can be used to structure polarization patterns. To do so, we take advantage of the flexoelectric effect and design splay structures that geometrically define the polarization direction. We demonstrate the creation of periodic polarization structures and the possibility of guiding polarization by embedding splay structures in uniform backgrounds. The demonstrated capabilities of polarization patterning, open a promising new route for the design of ferroelectric nematic based photonic structures and their exploitation.**

The recent experimental realization of ferroelectricity in a 3D fluid satisfies a century of speculation, and is anticipated to be widely exploited in emerging technologies, with significant societal impacts. Known for years and broadly exploited in modern display technologies, the standard nematic liquid crystal (NLC) state is uniaxial and non-polar. In 2017 two independent groups reported the occurrence of several N phases in two materials, RM734[1] and DIO[2]. The lower temperature phase of DIO was shown to exhibit ferroelectric order[2]. Subsequent works have identified the novel N phase in both compounds as a ferroelectric nematic ($N_F$) phase, in which inversion symmetry is broken leading to large spontaneous electric polarization ($P\sim6\,\mu C/cm^2$ for RM734[3] and $P\sim5\,\mu C/cm^2$ for DIO [4]). It has been shown that, the growth of polar order and splay deformation are connected in RM734, i.e. the N-$N_F$ transition is a ferroelectric-ferroelastic phase transition in which the growth of ferroelectric order is accompanied by the softening of the splay orientational elastic constant[5,6].

The technological relevance of NLCs relies on the multifaceted combination of optical anisotropy, responsiveness to electric fields and standardized alignment control via surface treatment of the confining media. Ferroelectric NLCs (FNLCs) have additional applicative potentialities owing to their unique combination of fluidity and spontaneous polarization[4,7,8], giant dielectric permittivity[7,9] and second order non-linear optical properties[6,8,10–13] enabled by the lack of inversion symmetry. The exploitation of the latter is at present restricted by the limited ability to control and shape the polarization direction via surface boundary conditions. Besides orientational coupling, FNLCs are guided by polar coupling constrains[10,14,15], affected by surface charges around defects and highly restricted by the large depolarization field entailed by the large magnitude of the polarization vector **P**.

In analogy to piezoelectricity, in which strain induces polarization, splay and bend orientational field deformations in NLCs can cause electric polarization of the medium, although in non-polar NLCs such effects are small[16]. Here we experimentally demonstrate that in FNLCs the flexoelectric coupling between deformation and polarization is strong and can be used to control the polarization direction. To demonstrate this, we design a series of polarization structures realized via periodic splay photoalignment-enabled patterns. We explore the differences between DIO and RM734 FNLC materials and investigate the stability of the observed structures via a simple model including elastic and electrostatic torques.

## Monodomain structures and alignment quality

RM734 and DIO (phase sequence and structure in Supplementary Fig.1) are filled at a temperature of the N phase into glass cells with patterned photoalignment and cell thickness (gap) $d$ (see Methods and Supplementary Fig.2). Each cell contains an array of different photopatterned 1.3 x 0.73 mm$^2$ rectangles (See Methods). Patterning via photoalignment prescribes at the cell surfaces anisotropic nonpolar in-plane alignment of the nematic director **n** (i.e. the direction of the average orientation of the molecules in the nematic phase) without an out-of-plane tilt. Two distinct electrode configurations were employed: (i) cells with top and bottom surfaces with uniform ITO electrodes and (ii) cells with one uniform ITO surface and one surface patterned with interdigitated electrodes of 100 μm width and 1 mm gap. For both materials, the transition to the $N_F$ phase on cooling materializes in a two-step process, in which the phase transition is followed by a structural relaxation (**n** structure changes across the confining cell thickness often involving twisting[10,14]), which is conditioned by the surface anchoring. In the photopatterned cells, such structural relaxation occurs very close to the N-$N_F$ temperature for RM734, and the intermediate structure cannot systematically be stabilized. However, in the case of DIO we have observed a temperature range of ~ 5-6 °C below the $N_S$-$N_F$ transition at which the undistorted state is stabilized. This structural relaxation occurs via the development of structured domain walls (cell background in Supplementary Fig.4), and we will refer to the director structures as *before* and *after* structural relaxation.

We first evaluated orientational and polar surface anchoring of photoalignment in the $N_F$ phase for uniform patterns (Fig.1). Remarkably, despite the lower anchoring strength expected for photoalignment, azimuthal surface anisotropy results in large monodomain patterns for both materials and those cells with uniform ITO at both surfaces (Fig.1a-c&e). While the structural relaxation takes place in the surrounding unaligned areas, it skips the uniform patterns (Supplementary Fig.4) which remain untwisted. Orientation is maintained for increasing cell gaps d= 3, 5 and 8 μm (Supplementary Fig.5). We investigated the homogeneity of the polarization by Second Harmonic Generation microscopy (SHG-M, see Methods) and interferometry microscopy (SHG-I). Interferometry conditions were set according to the main SHG signal contribution (Supplementary Fig.6). The SHG interferogram (Fig.1g) for RM734 shows for different areas that phase of the SHG signal is the same, i.e. the polarization direction is preserved throughout the pattern.

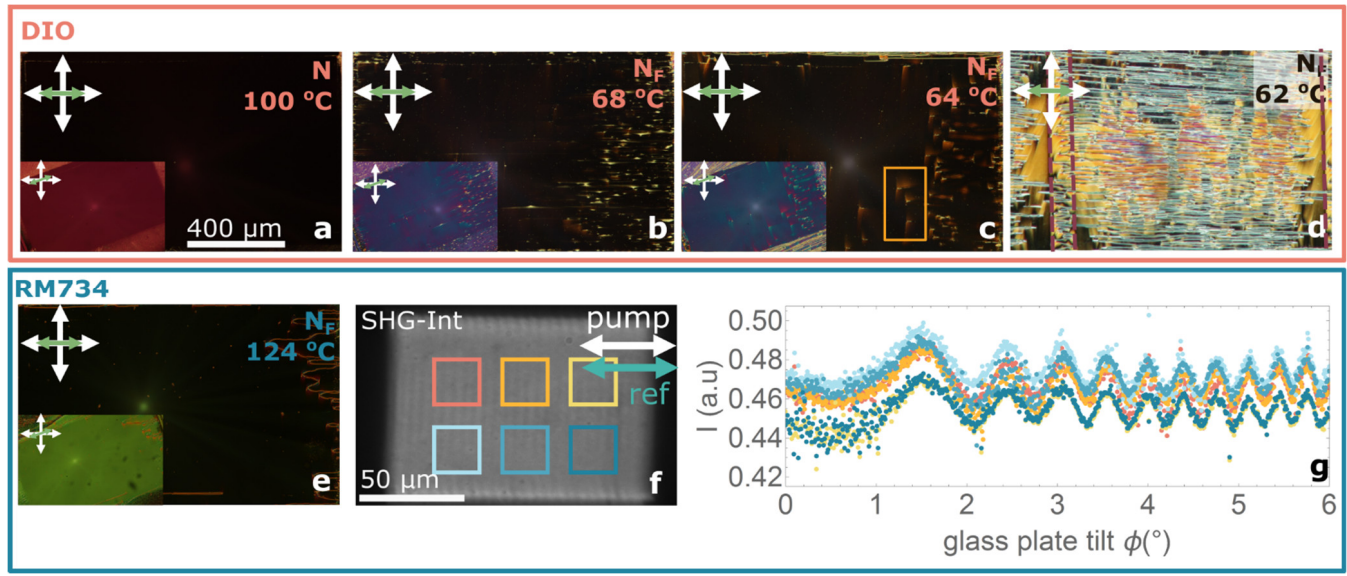

**Fig.1. Photopatterned millimeter range uniform polar domains. a-c** 1.3 x 0.7 mm photopatterned uniform domains in DIO in the N phase (a), in the $N_F$ phase before final structural relaxation (b) and in the $N_F$ phase after structural relaxation in a 3 μm thick cell (c). Main images show the patterned aligning direction of the investigated area (horizontal as indicated by bidirectional green arrow) aligned with the crossed polarizers (white arrows), while the inset show the same area under 20-degrees anticlockwise rotation. Microphotographs show a uniform structure with some defects (see Fig. 2) arising from the unpatterned edges of the structured area. **d** Same photopatterned structure in a region of an cell where one of the surfaces has interdigitated electrodes shows that the absence of conductive surfaces leads to an uncontrolled division in ferroelectric domains due to uncompensated charges and polarization distortions due to local impurities. Dashed maroon lines in d mark the position of the electrodes. In the broad region between them, only one of the confining cell surfaces has ITO electrode. **e** Observations of the same photopatterned structure filled with RM734. **f** SHG-I image of the RM734 uniform structure and **g** SHG interferogram corresponding to the highlighted areas in **f**. Measurements in the different areas show the same phase, indicating that the polarization direction is the same throughout the uniform pattern.

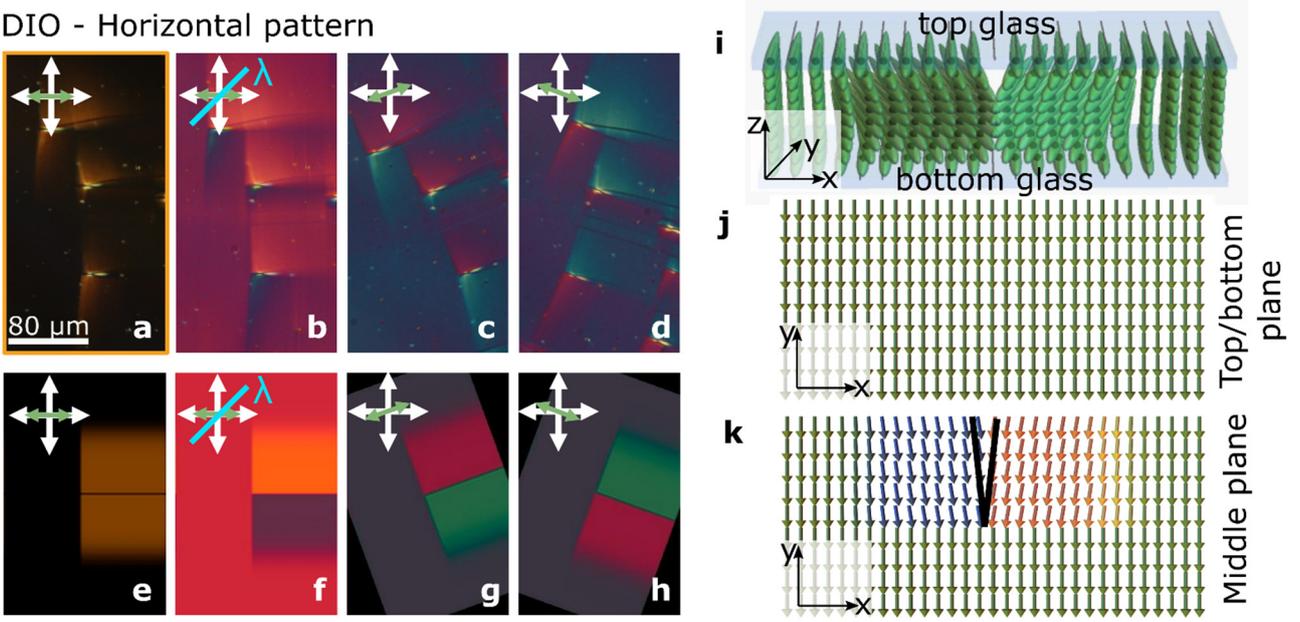

**Fig.2. Dragonfly-like splay deformation due to charged defect lines. a-d** Zoom-in POM images of the area highlighted in Fig.1c at different geometries **a** extinction position, **b** with full wave plate at 45° and **c&d** with the sample rotated in opposite directions. Images show a distortion of the uniform structure around spike-shaped defect lines. Crossed bidirectional white arrows indicate the direction of the polarizer and analyzer while the green arrow indicates the orientation of the photopatterned direction. **e-h** Dtmm simulations for the POM geometries in a-d, considering the structure described in i-k. **i** 3D sketch of the director and polarization arrangement around a defect line in the uniformly patterned area. Here, light blue rectangles represent the cell surface glass plates. **j** and **k** show the corresponding sketch in the cell surface plane (j) and in the middle of the cell plane (k) respectively. Structure twists across the cell thickness to accommodate the splay around the defect line. This splay director distortion reflects the electric charge of the defect line.

Comparison with the pattern in Fig.1d, evidences the important role of the electrodes. In the case of charged surface impurities, floating conductive electrodes compensate the charges, effectively equalizing the potential across the cell and hindering polarization distortions. Conversely, in Fig.1d, the absence of a uniform electrode in one of the surfaces, leads to uncompensated charges and polarization distortions that lead to the subdivision in numerous domains. On top and around the electrode, subdivision in domains is partially prevented.

Occasionally defects, formed at the transition, are observed at the pattern edge before wall propagation. After the structural relaxation they lead to domain walls surrounded by uniform alignment in RM734, but interestingly, in DIO to spike-like disclination lines (Fig. 2 shows the zoom-in of Fig.1c framed area). Such lines unequivocally create a characteristic optical distortion around them that looks like a dragonfly, with long wings extending transversally to the defect and sharply ending at the defect bending-tip. We deduced the underlying director structure by combining polarizing optical microscopy (POM) examinations (Fig.2a-d) and diffractive transfer matrix method (dtmm[17]) optical simulations (Fig.2e-h). At the surfaces, the director **n** is determined by the anchoring and it twists towards the middle of the cell to accommodate an in-plane splay as shown in Fig.2i-k. Such splay distortion evidences that topological defect lines in the $N_F$ in this case carry electric charge.

## Control of polarization direction via flexoelectric coupling

Considering the important role of splay deformations in the preceding nonpolar nematic phase[5,6] and stimulated by the quality of the observed alignment, we fabricated a series of periodic splay structures (Fig.3). The surface anchoring is shaped as $\mathbf{n}_s = (\sin(\vartheta_{surf}), \cos(\vartheta_{surf}))$, $\vartheta_{surf} = \vartheta_0 \sin(2\pi x/P)$, with $\vartheta_0$ ranging between 20 and 60 degrees and the period $P = 2\pi/k$ ranging between 20 and 60 μm (Supplementary Fig.7). The maximum splay curvature $k\vartheta_0$ of these patterns is then between 0.04 and 0.3 μm$^{-1}$. In the $N_F$ phase, the photoalignment results in large periodic arrays of lines with controlled alignment (Fig.3a) and remarkably, this is so regardless of the ITO patterning conditions (Supplementary Fig.8), which highlights the key role of splay $-\nabla \cdot \mathbf{P}$ "bound" charges in shaping the domain formation. We first focus on those patterns filled with DIO at a temperature of the $N_F$ phase, before wall propagation. Careful inspection of POM textures during the $N_S$-$N_F$ transition (Supplementary Fig.8 and Video 1) shows, immediately after the transition, the formation of well-defined disclination lines running up-down the pattern where the splay changes sign (Fig.3.b). Combining POM observations together with dtmm simulations, we established the approximate $\mathbf{n}(\vec{r})$ structure (Fig.3.d-f. and Supplementary Fig.11 to Fig.15). While at both surfaces $\mathbf{n}(\vec{r})$ splays to follow the anchoring, towards the cell centre it twists to recover a uniform structure (Fig.3.d-f). Such "unsplay" can be described by $\vartheta(z) = \vartheta_{surf} e^{(2z^2/d^2 - 1/2)/\xi}$ with $\xi = 0.2$ (see Supplementary Fig.11).

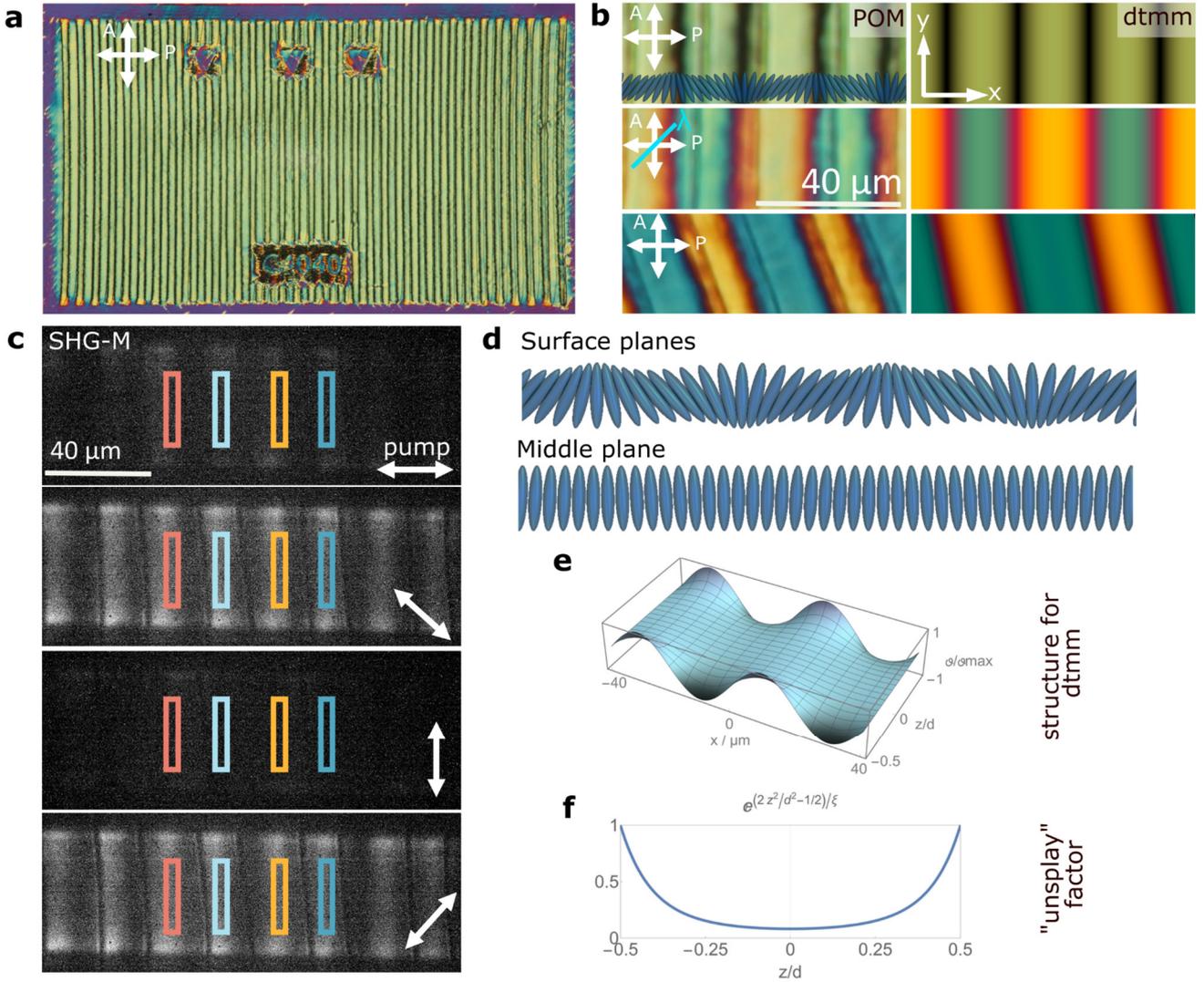

**Fig. 3. Periodic splay patterns. a** Overview of the Splay-4040 periodic splay patterned area in a 3 μm cell filled with DIO, where the azimuthal photopatterned angle varies periodically as $\vartheta_{surf} = \vartheta_0 \sin(2\pi x/P)$ with maximum angle $\vartheta_0 = 40°$ and period 40 $\mu m$ (splay curvature 0.1 μm$^{-1}$). The three top squares contain arrows pointing into lines in which **n** is vertical and the bottom rectangle provides the structure label. **b** Zoom-in on the structure showing the POM observations (left) along with the corresponding dtmm simulations of the transmitted spectra (right) at different geometries (splay lines parallel to analyser, with a full lambda plate at 45 degrees and sample rotated anticlockwise for 20 degrees) taking into account the structure described through d-f. **c** SHG-M images of a Splay-P40A40 pattern at different incoming laser polarizations, indicated by the bidirectional white arrow. **d** Schematic representation of the director orientation at the surface and in the middle plane of the cell used for dtmm simulations in **b**. **e** Azimuthal angle $\vartheta(z,x)$ across 2 structure periods (x between -40 and 40 μm) and the confining cell thickness (z from -0.5 d to 0.5 d). **f** Profile for the reduction of the azimuthal angle across the cell thickness.

SHG-M shows a rather uniform SHG signal within two consecutive disclination lines (Fig.3c), in agreement with the described tendency towards uniform **n** orientation in the middle plane. The SHG signal (Fig.3.g) dependency on incoming laser polarization directed is in agreement with the observations in Supplementary Fig.6 and determines the geometry for SHG-I measurements. Notably, disclination lines remain SHG inactive at any incoming polarization of the pump laser.

POM and SHG-M observations cannot distinguish the sign of **P** Structures with **P** alternating with the splay direction or remaining the same produce equivalent POM and SHG-M observations. As the phase of the generated SHG light depends on the direction of **P**, regions with opposite polarity can be distinguished by changing the phase of a reference SHG signal. For DIO, the reference BBO crystal was placed so that the SHG$_{ref}$ polarization lies perpendicularly to the periodic stripes. SHG-I measurements (Fig.4.a-b) demonstrate that contiguous areas with opposite splay and separated by disclination lines, have opposite direction of **P**, as the generated signal has a phase shift of 180° between them. These observations experimentally demonstrate how to exploit the flexoelectric coupling between polarization and orientational deformation for the design of polarization structures. Splay and bend director distortions in LCs can cause electric polarization of the medium even in non-polar NLCs. This effect, coined flexoelectric effect by de Gennes[16] and originally described by R.B. Meyer,[18] can be illustrated considering a nematic phase formed by pear-shaped or bent-shaped molecules with dipole moment along the long and short axis, respectively. In the first case, for a splay deformation, the minimization of the excluded volume results in the appearance of a macroscopic electric polarization proportional with $\mathbf{n}(\nabla \cdot \mathbf{n})$. Correspondingly, in the N$_F$ phase with spontaneous polarization **P**, flexoelectric coupling is also allowed by the symmetry and the polar order couples with director deformations, implying that one direction of polarization is more favourable for a given splay deformation (Supplementary Fig.16). Indeed, this effect spontaneously generates the periodic flippling of polarization in our studied periodic splay structures. Alternatively, when the system is described with only one order parameter (**P**), the effect is called flexodipolar coupling, i.e. coupling of **P** with deformations of **P**[19].

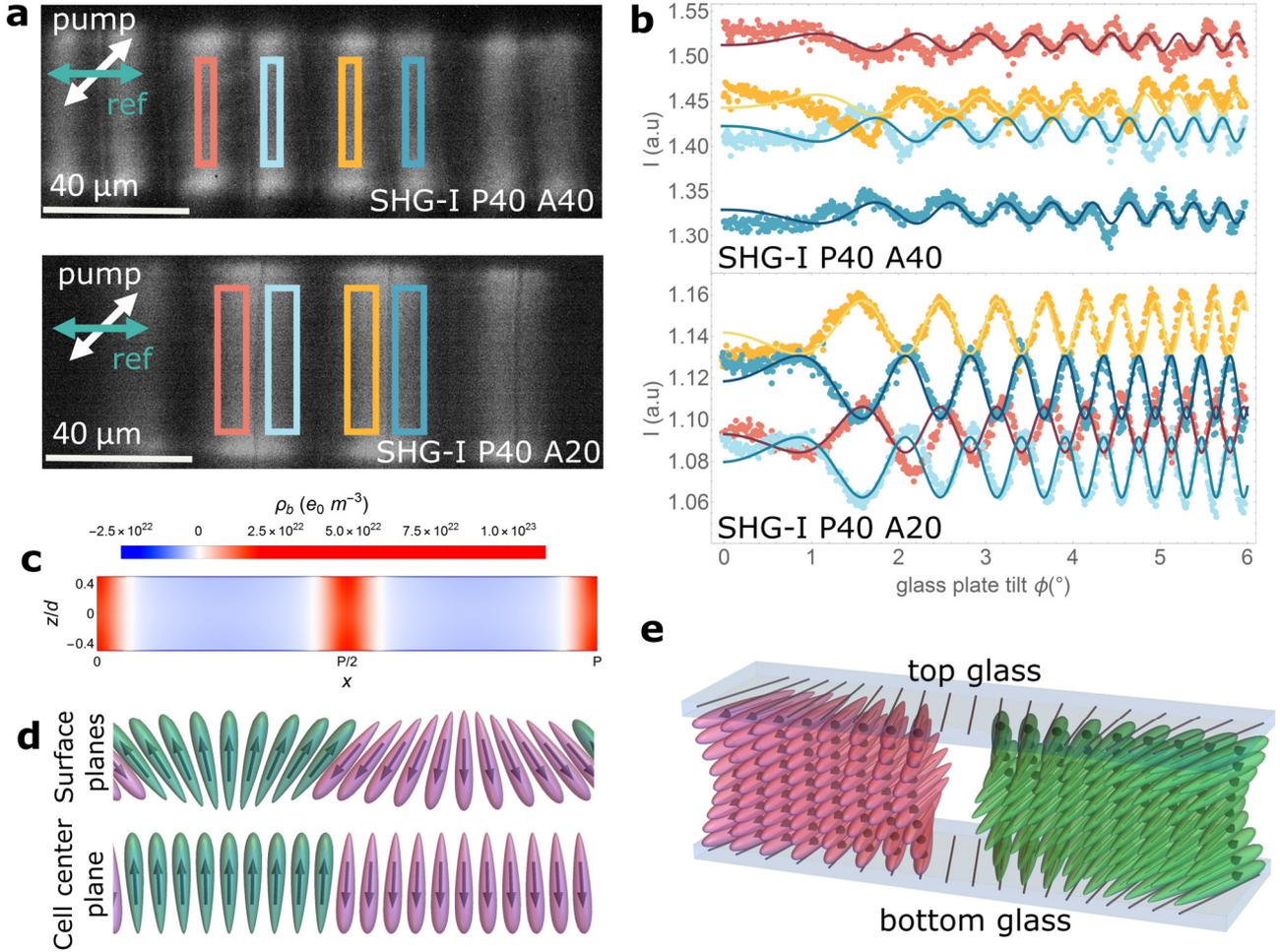

**Fig. 4. Polarization patterning proved by SHG interferometry in periodic splay photopatterned structures. a** SHG interferometry images of the periodic-Splay structures Splay-P40A40 and Splay-P40A20, with period 40 μm and maximum splay angle 40 and 20 degrees respectively (splay curvatures 0.1 and 0.05 μm$^{-1}$). Incoming polarization of the pump laser was at 45 degrees as indicated by the white arrow, the reference signal was horizontal, and the signal was collected after a horizontal analyzer. **b** SHG interferograms corresponding to the highlighted areas in (a) with matching colors. The SHG phase for two neighboring splay domains is opposite, i.e. polarization lies in opposite directions for splay regions of opposite sign. **c** Calculated charge distribution $-\nabla \cdot \mathbf{P}$ across the cell thickness $d$, assuming the relaxed structure initiated from that described in Fig. 3 for two consecutive splay lines as depicted in (d). Relaxed structure is shown in Fig.5.a&b and calculation details can be found in Supplementary Note VI. **d&e** Schematic representation of the surface and cell center planes (d) and 3D sketch (e) showing the flexoelectric coupling between the splay deformation and the polarization direction.

Splay deformation of **P** entails bound charges that can be calculated as $-\nabla \cdot \mathbf{P}$ (Fig.4c) that generate a depolarization field, which tends to suppress the splay, i.e. while splay is sustained in the surfaces due to the anchoring restrictions, the depolarization field suppresses it in the bulk as deduced via dtmm. This can be corroborated by means of a simple model to which we shall return in the next section.

Disclination lines (Fig.3&4) separate regions with opposite **P** directions. We noticed that, although always appearing at the edge of the splay pattern, occasionally the disclination lines can move to the centre of the splay region (i.e. they are not pinned to the surface), resulting in a slightly shifted structure (Supplementary Fig.10). Taking into account that, due to the polarity of the phase, their topological charge must be integer and that, to avoid electrical charge in the core, twist deformations are the most favourable, we envision the possible structure of such lines as twist-like disclinations with topological charge ±1 as sketched in Supplementary Fig.25. These structures are electrically charged, and thus, any splay structure asymmetry will cause their shift towards the centre of the splay, which carries opposite electric charge (Fig4.c).

On further cooling, the structural relaxation can also be detected in the photopatterned splay structure, characterized by the formation of domain walls, which initiate at the edges of the structure and propagate along the disclination lines (Supplementary Fig.8 and Video 1). The optical transmission texture in DIO remains unaltered, with a slight distortion around the newly formed domain walls, in line with the above-described observations of the dragonfly-shaped distortions. In the case of RM734, disclination lines at the edge of the splay regions are also observed right at the transition (Supplementary Fig.9). However, the subsequent structural relaxation takes place immediately after the N-$N_F$ transition, and although in the 3 μm cells alignment is reasonably preserved, in thicker (5 and 8 μm) cells a strong tendency to form $\pi$-twisted domains has been detected.

## Modelling of the periodic structure

To assess the stability of the periodic splay structures deduced from POM we use a simplified model incorporating the depolarization field, whose effect in the $N_F$ case is relevant in defining the director structure (Supplementary Note VII). First, we calculate local fields for the above-described structures and subsequently, we employ the Euler-Lagrange formalism to compute the relaxed structure. For this, a simple model is considered, in which Frank-Oseen elastic torques[20] are counteracted by electric torques.

In general, ferroelectric nematic liquid crystals can be described by two coupled order parameters, a nematic quadrupolar, i.e., tensor $\mathbf{Q} = S(\mathbf{n} \otimes \mathbf{n})$, and the electric polarization vector $\mathbf{P}$[21]. Here, $S$ is the scalar order parameter and $\mathbf{n}$ the director[16]. Here, we made the following assumptions: (i) $S$ is constant, so the nematic order can be described only by $\mathbf{n}$; (ii) $\mathbf{P} = \mathbf{P}_s + \boldsymbol{\varepsilon}_0(\boldsymbol{\varepsilon} - \mathbf{I})\mathbf{E}$, where $\mathbf{P}_s = P_0\mathbf{n}$; (iii) $P_0$ is constant; (iv) dielectric tensor is isotropic, $\boldsymbol{\varepsilon} = \varepsilon\mathbf{I}$; and (v) the orientation of the director at the surface is the same as prescribed by photopatterning. While the effective value of dielectric constant measured by dielectric spectroscopy is large, i.e. of the order of 10000 (which mainly comes from the reorientation of $\mathbf{P}_s$)[21], the value of $\varepsilon$ as defined above only accounts for induced polarization and is expected to be of the order 100 – 1000.

By minimization of a Landau-de Gennes type free energy functional, we calculated stable structures for the splay patterns. The functional additionally includes the electrostatic potential[22] to account for bound charges due to $-\nabla \cdot \mathbf{P}$ and free ions. Assuming that the dynamics of $\mathbf{n}$ (and $\mathbf{P}_s$) are much slower than that of free charges, then, during $\mathbf{n}$ relaxation, the local field $\mathbf{E}$ is given by the Poisson-

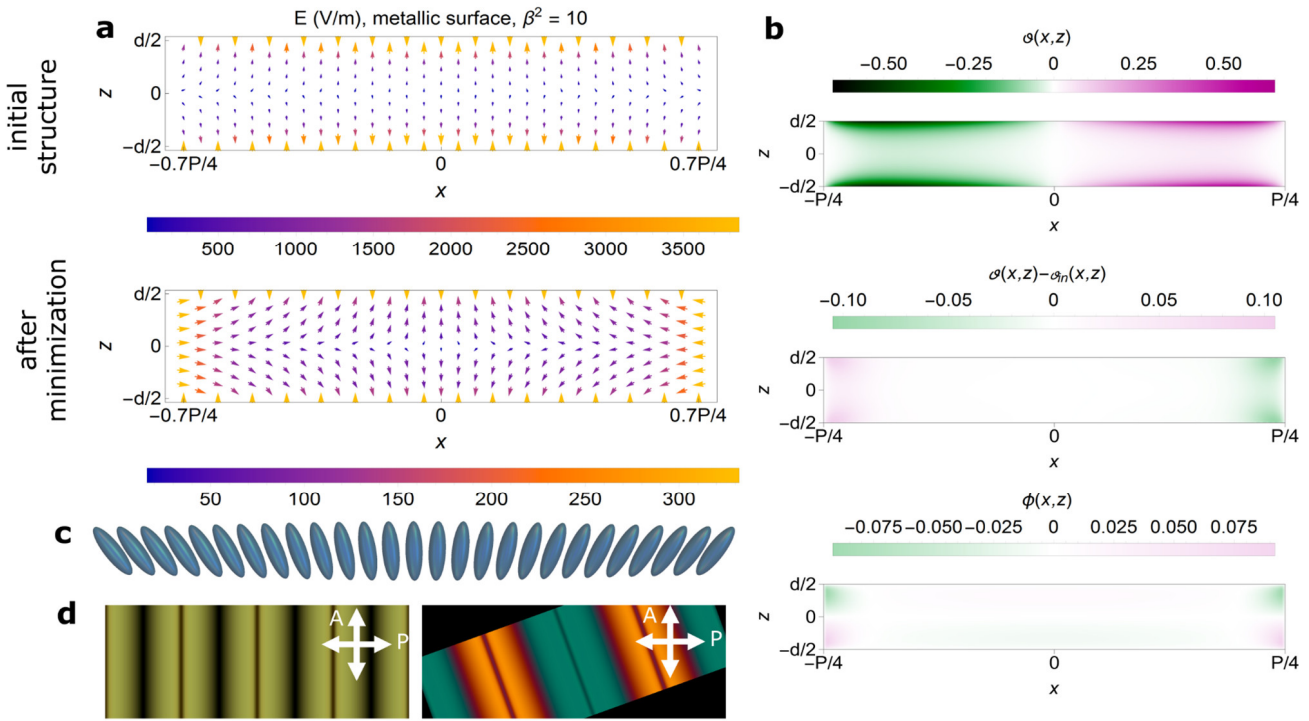

**Fig. 5. Assessment of structure stability via electrostatic calculations. a** Comparison of the local field for initial structure (top) and the structure after relaxation (bottom) in the central region ($-0.7\ P/4 < x < 0.7\ P/4$) as depicted in **c** for $\beta^2 = 10$ and $\varepsilon = 100$. The x and z coordinates are given in units of $\xi_b = 63$ nm. **b** Comparison of the structure before and after relaxation in the central region for $\beta^2 = 10$ and $\varepsilon = 100$. The initial structure is given by the angles $\vartheta_{in}(x,z)$ and $\phi_{in}(x,z) = 0$ as described in Eq.3 and Eq.4 of the Supplementary Note VII. Angles in the plot are given in radians. **c** Sketch of the splay section corresponding to the local fields shown in (a). **d** Dtmm simulations corresponding to the relaxed structure defined by the angles $\vartheta(x,z)$ and $\phi(x,z)$ shown in (b) as observed between crossed polarizers (crossed white arrows) when oriented parallel to them or with the sample rotated 20 degrees.

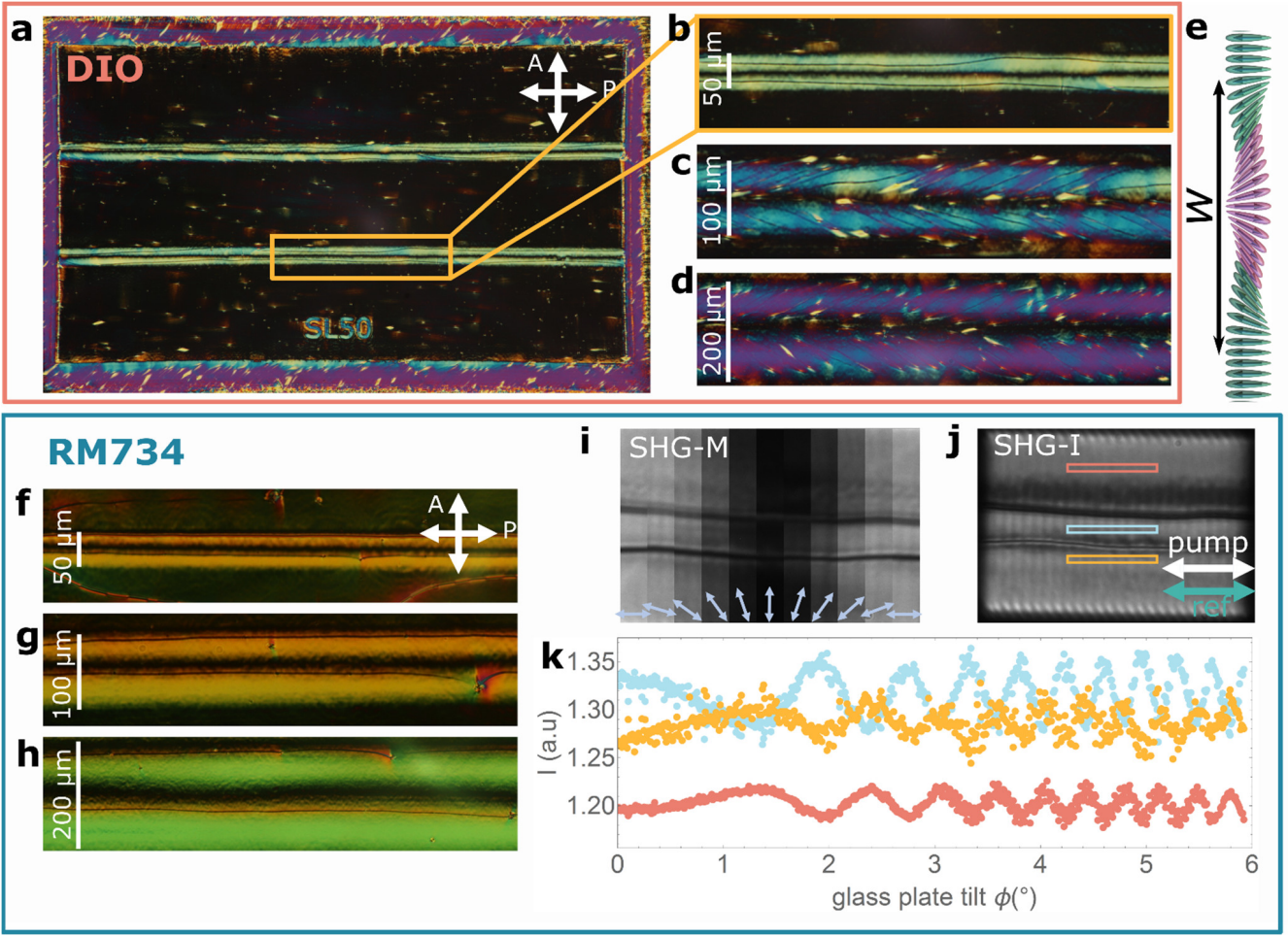

**Fig.6. Single splay lines embedded in uniform background for guiding polarization. a** POM overview of the photopatterned 1.3 x 0.7 mm² structure filled with DIO. **b** Zoom in of the area highlighted showing the formation of two disclination lines along the splay edges of the single line embedded in a uniform background. **c&d** Patterns with reduced splay curvature ((c) 0.05 and (d) 0.025 μm⁻¹) in DIO fail to produce controlled polarization domains and result in the appearance of multiple disclination lines oriented following the surface pattern. **e** Scheme of the photopatterned structure in which a splay line of width W/2 is embedded. **f-h** Similar splay lines for a 3 μm cell filled with RM734. In the case of RM734, lower splay also results in controlled patterning of polarization directions. **i** shows a reconstruction of SHG-M images taken at different incoming pump laser polarizations. **j** SHG-I image of the same area depicted in (i). For SHG-I the incoming pump polarization and SHG reference signal were selected along the patterned lines and the intensity was recorded with a horizontal analyzer. **k** SHG interferogram for the three areas highlighted in (j) showing opposite polarization direction in the area between both disclination lines (one splay direction) and outside them (opposite splay direction).

Boltzmann equation and the relaxation method can be used to minimize the relevant part of the Landau-de Gennes free energy:

$$F_\mathbf{n} = \int \left(\tfrac{1}{2}K_1|\mathbf{S}-\mathbf{S}_0|^2 + \tfrac{1}{2}K_2 Tw^2 + \tfrac{1}{2}K_3|\mathbf{B}|^2 - \tfrac{1}{2}P_0\mathbf{n}\cdot\mathbf{E}\right) dV$$

Here, $K_i$ ($i$ = 1,2,3) are splay, twist, and bend elastic constants with corresponding deformations $\mathbf{S} = \mathbf{n}\nabla\cdot\mathbf{n}$, $Tw = \mathbf{n}\cdot(\nabla\times\mathbf{n})$, $\mathbf{B} = \mathbf{n}\times(\nabla\times\mathbf{n})$, and $\mathbf{E} = -\nabla\Phi$. The flexoelectric coupling is added to the first term, where $\mathbf{S}_0 = \gamma\mathbf{P}/K_1$. The sign of the flexoelectric coefficient $\gamma$ determines which direction of $\mathbf{P}$ is favourable when a splay deformation $\mathbf{S}$ is present. The ideal splay curvature which would minimize the splay elastic energy is $\mathbf{n}\cdot\mathbf{S}_0$. The relaxation steps were performed with respect to $\vartheta(x,z)$ and $\varphi(x,z)$. At each step, $\mathbf{E}$ was recalculated using the linearized Poisson-Boltzmann equation (Supplementary Note VII, Eq. 7).

Mapping of the local fields for the initial and relaxed structures show large local fields and deformations at the edges of the splay pattern, where experimentally defects are observed (Fig.5a). For a large enough ions density ($\beta^2 \gtrsim 1$) the depolarization field originating from these regions is screened by the free ions, and thus has a negligible effect on the director structure in the centre of the splay line. The reduction of the local depolarization field for the structure after minimization is significant (Fig. 5a. for $\beta^2 = 10$, $\varepsilon = 100$, i.e. an screening length of 20 nm). However, the director structure changes are small, being negligible in the azimuthal direction (Fig.5.b) and more notable in the appearance of an out-of-plane splay deformation that causes the redistribution of bound and free charges resulting in the reduction of the local fields. Dtmm calculations using the relaxed structure show minor differences with respect to those using the initial structure (Fig.5.d).

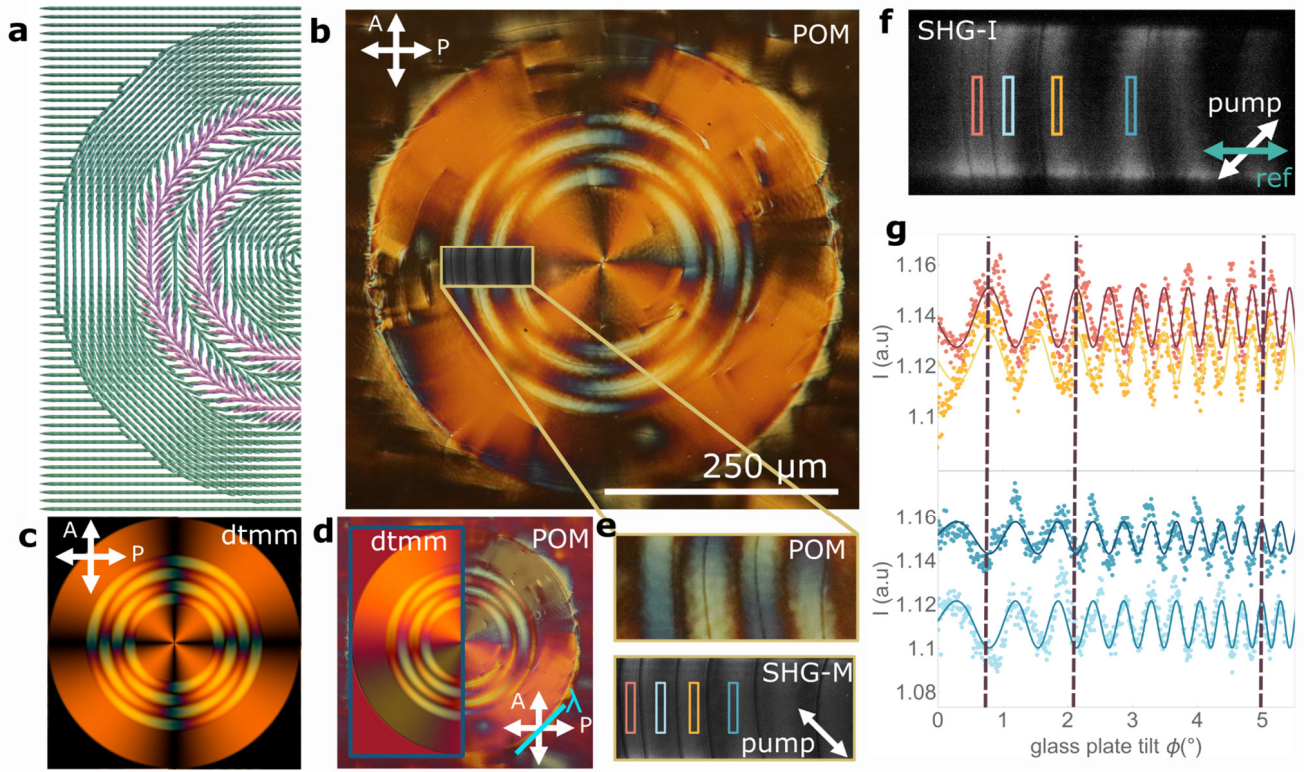

**Fig.7. Guiding of polarization with splay lines in a bend background. a.** Schematic of the photopatterned structure, in which two periods of a 4040 splay line are embedded tangentially into a pure bend circle deformation. Different arrowed colors highlight the alternation of regions with opposite polarization direction as deduced in (f-g) **b** POM image of the splay in bend lines in a 3 μm cell filled with DIO. Crossed white arrows indicate directions of the polarizers. Framed region shows the SHG-M image of that area. **c** Corresponding dtmm transmission spectra simulations considering a uniform bend structure in the outer and inner circles and a splay line deformation, with decreasing splay angle towards the center of the cell, equivalently constructed as the periodic splay lines considered in Fig.3 showing a nice correspondence with experimental observations. **d** Comparison of POM and dtmm observations of the same structure with the addition of a full lambda plate at 45 degrees with respect to the crossed polarizers, as indicated by the blue line. **e** Zoom image of the splayed region framed in (b) and corresponding SHG-M image with incoming laser polarization as marked by the white arrow. **f** SHG-I image of the same area as in (e). **g** SHG interferogram of the areas highlighted in f showing opposite phase of SHG signal across each of the disclination lines, indicating the reversal of polarization direction for opposite splay lines.

## Guiding polarization in uniform and bend environments

To assess the limiting splay curvatures needed to guide polarization, we first studied the inclusion of single splayed lines, with the same maximum splay angle (45 degrees) but differing width, into a uniform aligned pattern (Fig.6, $\vartheta_0 k =$ 0.1, 0.05 and 0.025 $\mu m^{-1}$). Remarkable differences can be observed between both materials. In DIO, the splay curvatures comparable to those in periodic patterns result in equivalent polarization guiding, while lower splay curvatures (<0.05) are not sufficient for producing controlled polarization structures as evidenced by the appearance of numerous disclination lines running along the splay photopatterned area (Fig.6b-d). However, for RM734 all investigated splay strengths show the characteristic formation of disclination lines and subsequent domain wall formation at the edges of the splay structure (Fig.6e-g). For both materials, the decrease of the splay curvature implies a decrease of the depolarization field, evidenced by the decrease of the "unsplay" of the structure (Supplementary Fig.17). SHG-M analysis in RM734 shows, in line with observations in the horizontal pattern (Supplementary Fig.6), a maximum of SHG signal for incoming polarization of excitation laser along the main director orientation. Correspondingly, SHG-I demonstrates the alternation of polarization direction across the domain walls (Fig.6i).

The observed differences between both materials should be attributed to distinct flexoelectric coupling properties and highlight the fact, that for the future development of controlled alignment techniques, the characteristics of each employed $N_F$ material need to be taken into account in order to find the optimum splay configuration.

To test the possibilities of guiding polarization direction via splay photopatterning through a deformed background, we designed a circular domain of radius 500 μm in which director lies tangentially, having pure bend deformation in the inner and outer rings, and an additionally overlaid set of circular splay lines ($\vartheta_0 = 45°$ and P=100 μm) in the radial area between 150 and 350 μm (Fig.7a). Optical investigation and comparison with simulated dtmm transmission spectra show that director orientation accurately follows the prescribed orientation throughout the cell thickness in the purely bend areas, while the degree of in-plane splay is decreased towards the cell centre similarly to what is observed in the periodic splay patterns (Fig.7b-d and Supplementary Fig.18). SHG inactive disclination lines are again evident, running tangentially along the splay lines (Fig.7.e) and separating regions of opposite polarization direction as demonstrated by SHG-I measurements (Fig.7f-g).

## Conclusions and outlook

We have demonstrated how to exploit flexoelectric coupling between polarization and splay director deformations in FNLCs to create controlled polarization direction patterns of interest for multiple non-linear photonic applications. Periodic splay patterns and single splay lines embedded in uniform or bent backgrounds, all show alternating polarization directions between regions of opposite splay, with the appearance of SHG inactive disclination lines at the edge of the inscribed splay structure. The depolarization field created by bound charges due to $-\nabla \cdot \mathbf{P}$ causes the escape of the surface prescribed splay structure towards a uniform orientation in the middle of the confining cell. The experimentally deduced structure is well explained by means of a simple model, which excludes the deformation around the charged defects. These results should serve as inspiration for further development of a model describing such defects in a general way, as done for the nonpolar nematic phase by Everts et al.[22]. The possibility to design custom polarization structures, patterning SHG signal and the prospect of steering and reconfiguring it by means of external electric fields is of great interest for applications in non-linear photonic devices, well beyond the classical multibillion technological implementations of nematic liquid crystals.

## Methods

### Materials

Synthesis of the liquid-crystalline material DIO (2,3',4',5'-tetrafluoro-[1,1'-biphenyl]-4-yl 2,6-difluoro-4-(5-propyl-1,3-dioxan-2-yl)benzoate) has been performed according to the description given in reference[12]. The molecular structure, together with the phase sequence are presented in Supplementary Fig.1. Due to the 1,3-dioxane unit, sample was always maintained below 120 °C to avoid changes in the molecular structure. On cooling from 120 °C, transition to Ns is observed at 83.9 °C and followed by the Ns-$N_F$ transition at 68.9 °C, as reported in reference 12. Subsequent crystallization temperature was observed at temperatures lower than 60 °C, temperature that differs depending on the cooling rate.

Description of the synthesis of the liquid-crystalline material RM734 can be found in reference[1]. The structure of RM734 (4-((4-nitrophenoxy)carbonyl)phenyl-2,4-dimethoxybenzoate) and the phase sequence are presented in Supplementary Fig.1. On heating, the crystalline phase melts directly into the nematic phase at 139.8 °C and transforms into the isotropic liquid phase at 187.9 °C. On cooling, the isotropic (I) to nematic (N) phase transition is followed by a nematic to ferroelectric nematic ($N_F$) transition at 132.7 °C, which crystallizes around 90 °C, temperature which can vary depending on the cooling rate.

### Photopatterning

The LC director at the substrate surface was defined through patterned photoalignment. After ozone-plasma treatment, a mixture of 0.2 wt% Brilliant Yellow (Sigma-Aldrich) in dimethylformamide (DMF, Sigma-Aldrich) is used for spin coating (3000 rpm during 30 s). The substrates with dimension one inch by one inch, are either homogeneously coated with ITO or have an ITO electrode pattern. The substrates are then placed onto a hotplate for 5 minutes at 90 °C. With two coated substrates a cell is fabricated, by placing glue and spherical glass beads near the edges of one substrate and placing the second substrate on top. When illuminated with linearly polarized light (UV or blue light), brilliant yellow provides in plane preferred orientation of nematic liquids perpendicular to the incident polarization. To achieve the desired patterns, the complete cell is then exposed to the illumination of a blue laser (Cobolt Twist, $\lambda$= 457 nm) that is modulated in polarization by an optical setup that contains a spatial light modulator (Holoeye Pluto 2). The pattern with 1920 by 1080 pixels is projected onto the cell, with the pixel voltage encoding the azimuthal angle of the linear polarization of the illumination. By moving the sample between subsequent illuminations, multiple patterns can be written in the same cell. For the patterns that provide a uniform alignment, an additional polarizer was inserted between the SLM and the LC cell.

### Polarizing optical microscope and Berreman calculus

Polarizing optical microscopy experiments were performed in a Nikon Eclipse microscope. Images and videos were recorded with a Canon EOS M200 camera. The sample was held in a heating stage (Instec HCS412W) together with a temperature controller (mK2000, Instec).

We used the "dtmm" open software package to calculate transmission spectra and colour rendering. The package uses the Berreman 4x4 matrix method to compute the transmission and reflection spectra. The microscope's lamp spectrum was measured and used as illuminant to calculate the transmission spectra. The latter are then converted to XYZ colour space using CIE 1931 colour matching function. Then the linear RGB colour from XYZ colour space was computed, as described in the sRGB standard IEC 61966-2-1:1999. Finally, in order to obtain the final nonlinear RGB colour values, suitable for display or print, we applied the sRGB transfer function (gamma curve) using D65 white point. By this procedure, the input daylight light source spectra is converted to a neutral grey colour in the case of uncrossed polarizers and no sample. To match the experimentally obtained images with the simulations, we performed in-camera white balance correction for daylight conditions to match the encoded D65 white point of the simulations allowing us to have a good agreement between the simulated and experimentally obtained images.

### Second Harmonic Generation microscopy and interferometry

Interferometric SHG imaging is performed using a custom-built sample-scanning microscope (Supplementary Fig.3). The laser source is a Erbium-doped fibre laser (C-Fiber A 780, MenloSystems) generating 785 nm, 95 fs pulses at a 100 MHz repetition rate. The average power was adjusted using an ND filter to 30 mW on the sample. The 1.2 mm beam diameter is expanded to 6.4 mm and focused to BBO crystal (Eksma Optics) to generate a reference SHG beam. In case of SHG Microscopy the BBO reference crystal is removed from the path. Polarization of the fundamental IR beam - marked with red - is perpendicular to the plane of the beam plane, whereas polarization of the reference SHG beam – marked with blue - is parallel to the plane. An off-axis parabolic mirror collimates the beams. Michelson interferometer is used for time compensation between the fundamental and the reference pulse.

The phase of the reference pulse is adjusted finely with a glass plate mounted on a motorized rotator. A half-waveplate for 800 nm rotates the polarization of the fundamental IR beam in the plane. This polarization was chosen horizontal for the measurements of RM734 and with an angle of 45° for DIO, according to results described in Supplementary Fig. 6. A motorized dual-wavelength half-waveplate adjusts the incident polarization on the sample and, jointly with the analyser in front of the camera, enables to perform polarization-resolved SHG. In the case of SHG-M images, analyser was removed to account for all contributions. A combination of galvo mirrors and and a long-working distance objective (Nikon CFI T Plan SLWD, NA 0.3) is used to scan the focused beam in the sample plane. The scanning frequencies are much higher (a few 100 Hz) than the imaging frame rate (a few Hz).

A long-working distance 20x objective (Nikon CFI T Plan SLWD, NA 0.3) collects the light coming from the sample. A set of 700 nm short-pass and 400 nm band-pass filter eliminates the fundamental IR light and any possible fluorescence signal. The SHG image is finally acquired using a high-performance CMOS camera (Grasshopper 3, Teledyne Flir) with a typical integration time of 250 ms.

## Data availability

The authors declare that the data supporting the findings of this study are available within the text of this manuscript, including the Methods and the Supplementary information. Source data are available from the corresponding author upon reasonable request.

## Acknowledgements

N.S, A.M, I.D-O., M.L, N.O and A.P acknowledge the support of the Slovenian Research Agency [grant numbers P1-0192, N1-0195 and PR-11214]. K.N and B.B would like to acknowledge the support of the Research Foundation – Flanders (FWO) through grant number G0C2121N. R.J.M acknowledges funding from UKRI via a Future Leaders Fellowship, grant no. MR/W006391/1. S.A. and M.H. acknowledge the National Key Research and Development Program of China (No. 2022YFA1405000), and the Recruitment Program of Guangdong (No. 2016ZT06C322).

## Author Contributions

N.S, A.M and K.N. designed and coordinated the work. N.S performed POM observations. N.S and A.M carried out dtmm simulations. A.P developed dtmm open software. M.L performed SHG-M and SHG-I experiments. N.O developed SHG-M and SHG-I setups. I.D-O assisted in the design and interpretation of SHG-M and SHG-I experiments. A.M. carried out electrostatic calculation. K.N and B.B. fabricated and characterized the photopatterned cells. R.J.M synthesized RM734. S.A. and M.H. provided DIO. N.S and A.M prepared the initial draft of the manuscript and all the authors made contributions to the final version.

## Competing interests

The authors declare no competing interests.

## Additional information.

Supplementary information is available for this paper.

# Supplementary Information

## Polarization patterning in ferroelectric nematic liquids

N. Sebastián[1], M. Lovšin[1], B. Berteloot[2], N. Osterman[1,3], A. Petelin[1,3], R. J. Mandle[4,5], S. Aya[6], M. Huang[6], I. Drevenšek-Olenik[1,3], K. Neyts[2], A. Mertelj[1]

1 Jožef Stefan Institute, Ljubljana, Slovenia
2 Liquid Crystals and Photonics Group, ELIS Department, Ghent University, Ghent, Belgium
3 University of Ljubljana, Faculty of Mathematics and Physics, Ljubljana, Slovenia
4 School of Physics and Astronomy, University of Leeds, Leeds, UK
5 School of Chemistry, University of Leeds, Leeds, UK
6 South China Advanced Institute for Soft Matter Science and Technology (AISMST), School of Emergent Soft Matter, South China University of Technology, Guangzhou, China
7 Guangdong Provincial Key Laboratory of Functional and Intelligent Hybrid Materials and Devices, South China University of Technology, Guangzhou, China

## Table of contents:

# Supplementary Note I – Supplementary Video descriptions

Supplementary Video 1 shows, for DIO, the transition between the antiferroelectric splay nematic phase and the ferroelectric nematic phase when confined in a photopatterned 3 µm cell in which the surfaces impose a periodic splayed structure with maximum splay angle of 40 degrees and splay period P= 40 µm. Pretransitional behaviour is characterized by a clearly visible stripe texture, with stripes following the photopatterned structure. At the transition disclination lines are formed uniformly along the pattern in those regions in which splay changes sign and progressively the texture between lines becomes uniform. Few degrees below the transition there is a final structural relaxation, characterized by the propagation of a final deformation along the disclination lines, forming domain walls. Such deformation involves charges that slightly reorient polarization around it, evidenced by the slight change in transmitted spectra. Snapshots together with the transition as observed with SHG-M are shown in Supplementary Fig. 8.

# Supplementary Note II – Materials and Methods

*DIO:* Iso  173  N  83.9  $N_S$  68.9  $N_F$  47  Cr

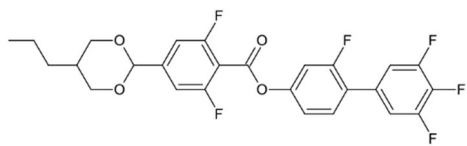

RM734: Iso 187 N 132.7 $N_F$ 90 Cr

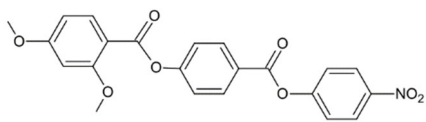

Supplementary Fig.1. Chemical structure of the two materials employed in this work, DIO and RM734, and their phase sequence on cooling[1,2]. While the latter shows on cooling a direct phase transition between the nonpolar nematic phase and the ferroelectric nematic phase, DIO exhibits three nematic phases; on cooling, a high temperature nonpolar nematic phase, followed by an antiferroelectric splay nematic phase and at lower temperatures the ferroelectric nematic phase focus of this work.

**Polarization patterning in ferroelectric nematic liquids**

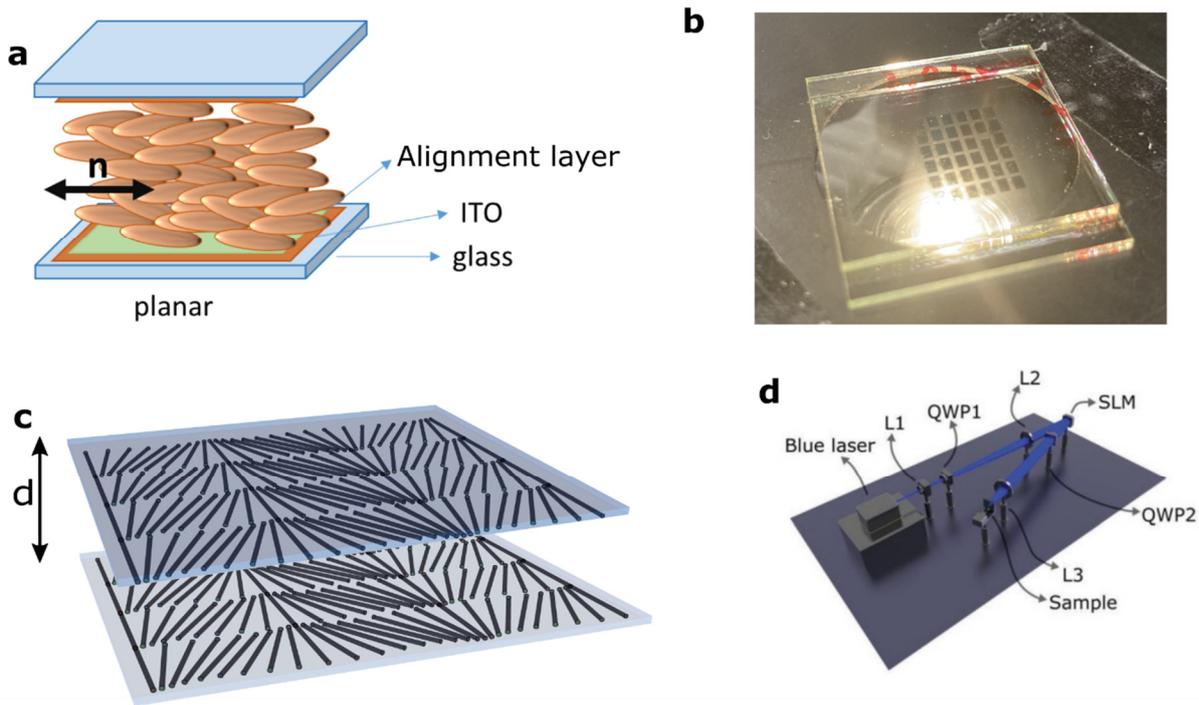

Supplementary Fig.2 Image illustrating typical liquid crystal cells and surface alignment via photopatterning. The described POM, SHG-M and SHG-I observations were done employing home-made glass cells, consisting of two transparent glass plates between which the liquid crystalline material is introduced by capillary action at a temperature of the nematic phase. In our case, both glass plates are lithographed with a transparent conductive layer of Indium Tin Oxide (ITO). Two different electrode geometries were employed, either uniform in both substrates, or interdigitated in one substrate and uniform in the other. In order to prescribe a defined in-plane orientation (planar alignment) of the nematic director, liquid crystal cells additionally incorporate an alignment layer. To achieve well-defined patterned directions of the molecular director at the cell interface, we exploit photoalignment technology, i.e. we can achieve non-uniform liquid crystalline alignment by use of a polarization-sensitive material, which will provide a preferred in-plane (planar) director alignment in a direction perpendicular to the incoming light polarization and illuminating a thin photoalignment layer spin-coated at the surfaces by pattered polarization illumination via a SLM device. a) Schematic representation of a liquid crystalline cell. b) Image of one of the photopatterned cells used in this study, in which the rectangular areas corresponding to different patterns can be clearly seen. c) The image illustrates as an example the alignment pattern at the bottom and top substrates for a periodic splay structure with maximum angle of 40 degrees. d) Scheme of the photopatterning setup, consisting on a blue laser, a spatial light modulator (SLM), three lenses and a quarter wave plates (QWPs)[3]. To achieve the patterned substrates, the cell's photoalignment layers are illuminated with linearly polarized light of which the polarization orientation varies spatially according to the desired orientation of the LC director at the substrates. To do so, the various orientations of linear polarization are generated using a spatial light modulator (SLM, Holoeye Pluto 2), with a resolution of 1920 by 1080 pixels and a pixel pitch of 8 µm. The SLM displays a grayscale image (= the photoalignment pattern) in which each grey level corresponds to a specific orientation of linear polarization. By means of a projection lens, the SLM displayed pattern was scaled down by a factor of 11.6. As light source a linearly polarized blue laser (Cobolt Twist, $\lambda$ = 457 nm) was used, that is first transformed into circularly polarized light and reflected via the SLM introducing a phase delay between the vertical and the horizontal field components. Such phase delay determines the azimuthal angle of the linearly polarized light that incises the photoalignment layers.

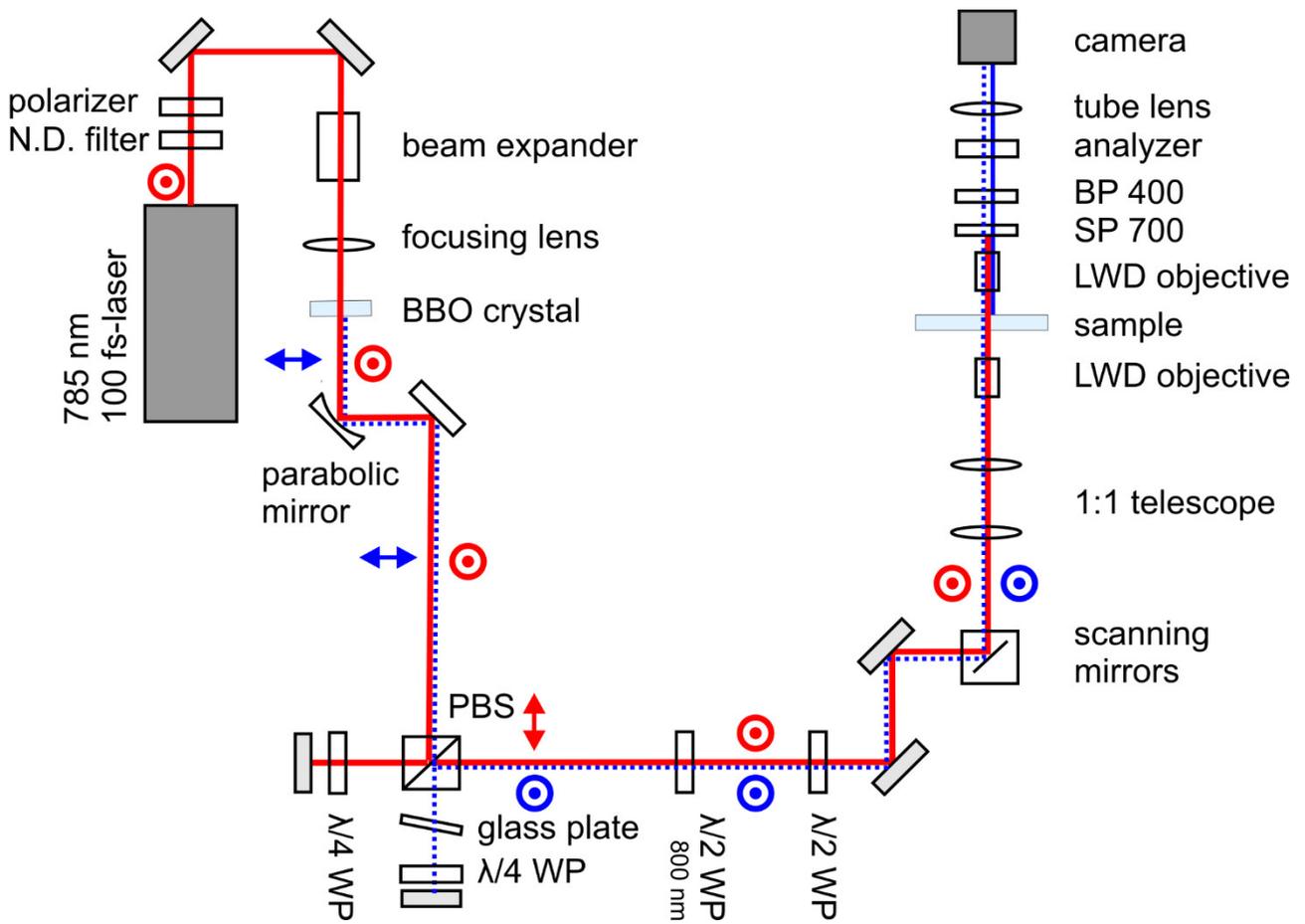

Supplementary Fig.3 Schematics of the Interferometric SHG imaging system as described in the main manuscript in Methods.

## Supplementary Note III – Uniform patterns

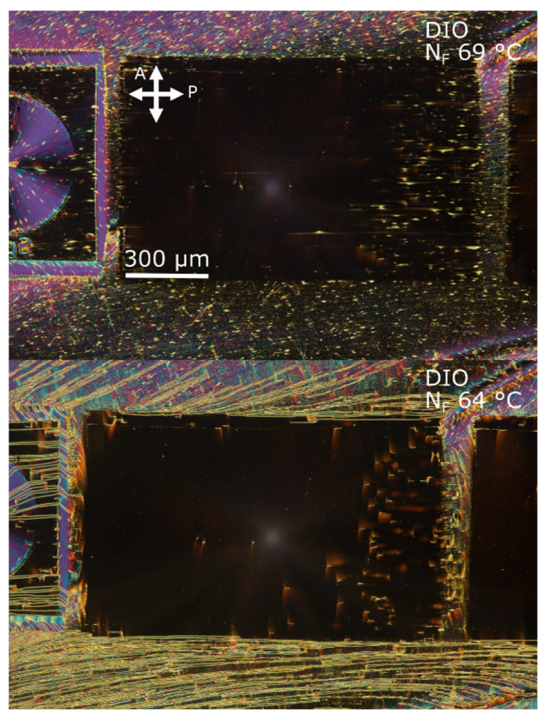

Supplementary Fig.4. Detail under crossed polarizers of the horizontal pattern and the surroundings for two temperatures in the ferroelectric nematic phase in DIO, before (69 °C) and after (64 °C) what is addressed in the manuscript as wall propagation. Both cell surfaces have a uniform conductive ITO layer. The appearance of domain walls is clearly observed in the surrounding unpatterned background and in the visible corner of the left pattern. The domain wall-free state can be stabilized for around 5 degrees under slow cooling. Cell thickness is 3 µm.

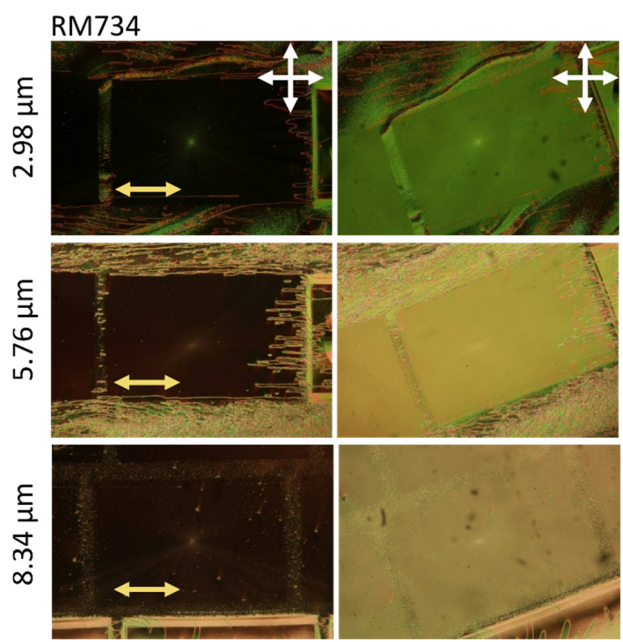

Supplementary Fig.5. Uniform photopatterned area in cells with different thicknesses d= 2.98, 5.76 and 8.34 µm filled with RM734 at a temperature in the $N_F$ phase (120 °C) showing uniform alignment throughout the 1.3 x 0.7 mm² patterned area.

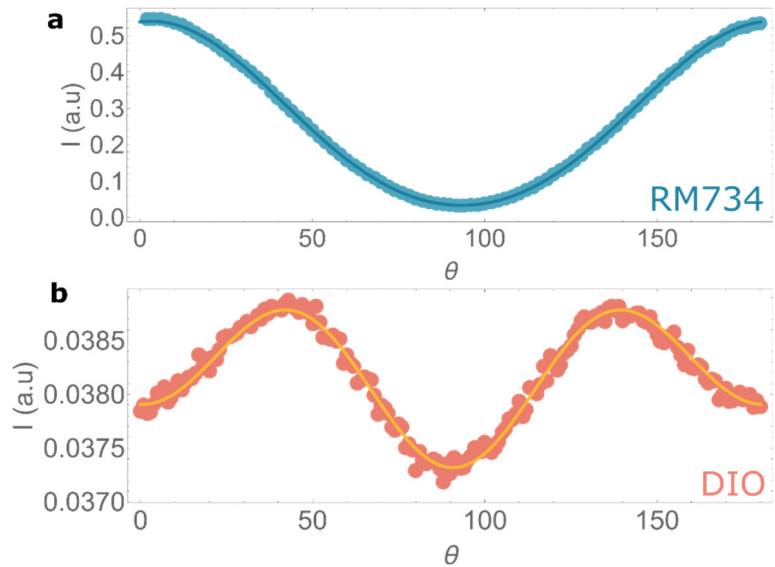

Supplementary Fig.6. Dependence of the recorded SHG intensity in horizontal patterns on the angle $\theta$, measured between the incoming laser polarization and the preferred photoalignment orientation. Results show for RM734 a maximum SHG signal for incoming polarization along the photoalignment orientation, while for DIO, the maximum is obtained at 45°. Such behaviour evidences that for RM734 the $\chi_{333}$ susceptibility coefficient is predominant, while for DIO the $\chi_{131}$ coefficient also contributes notably to the signal[4]. The results can be fitted by $A\cos^4(\theta) + B\cos^2(\theta)\sin^2(\theta) + C\sin^4(\theta) + D$.

## Supplementary Note IV – Periodic Splay patterns

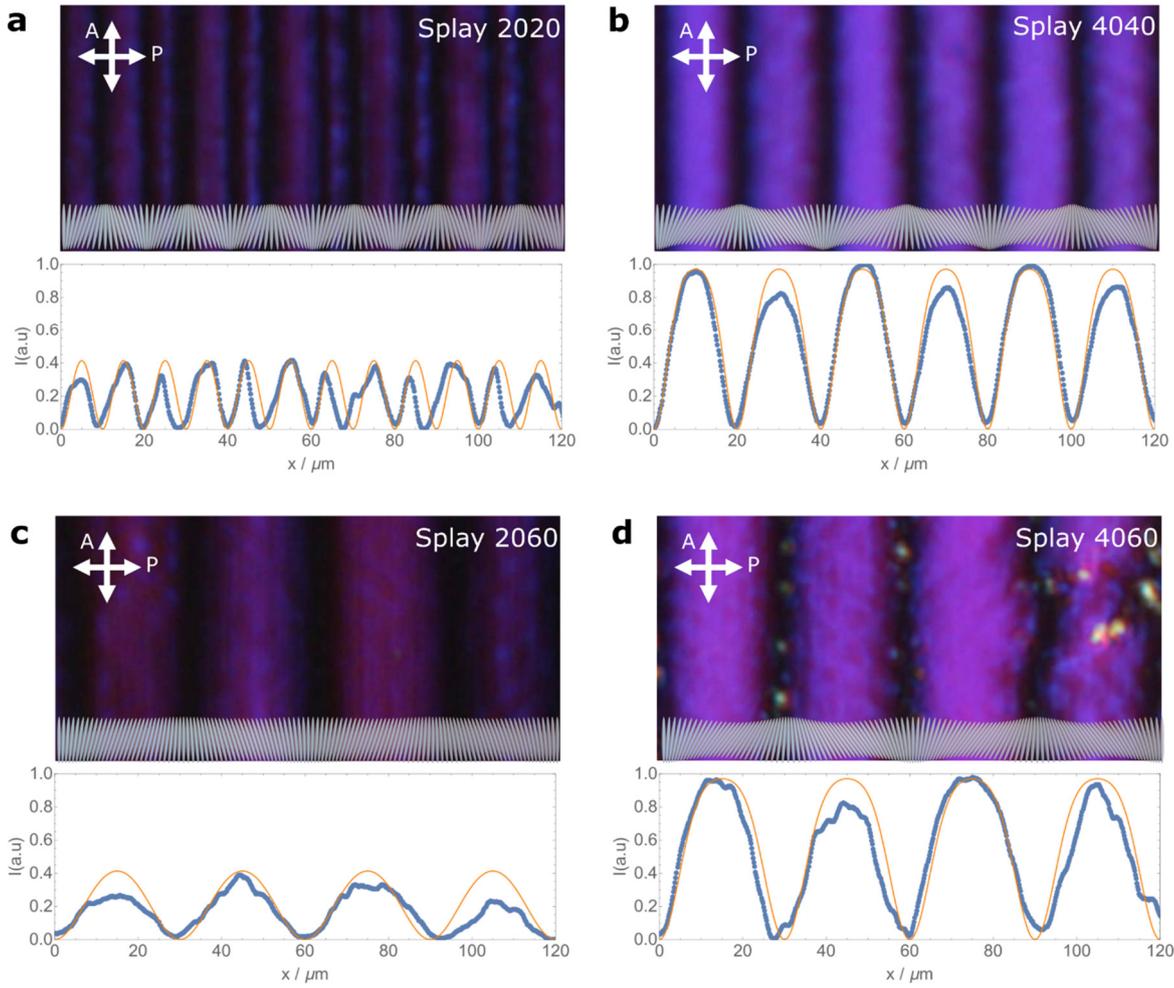

Supplementary Fig.7. Splay pattern profiles in the in the nematic phase 115 °C of DIO. POM images of 4 different one dimensional splay patterns in which the director deviates from the vertical direction by $\vartheta = \vartheta_0 \sin(2\pi x/P)$, where x corresponds to the horizontal direction, $\vartheta_0$ to the amplitude and $P$ to the period of the pattern are: a) $\vartheta_0 = 20°, P = 20\ \mu m$, b) $\vartheta_0 = 40°, P = 40\ \mu m$, c) $\vartheta_0 = 20°, P = 60\ \mu m$ and d) $\vartheta_0 = 40°, P = 60\ \mu m$ and illustrated by the overlaying sketches. Intensity plots show the comparison of the normalized measured intensities across the horizontal direction calculated from the POM images (blue circles) and the theoretical expected intensity profile $I = I_0 \sin(2\phi)^2$ where $\phi$ is the angle between the prescribed director orientation and the crossed polarizers (orange line).

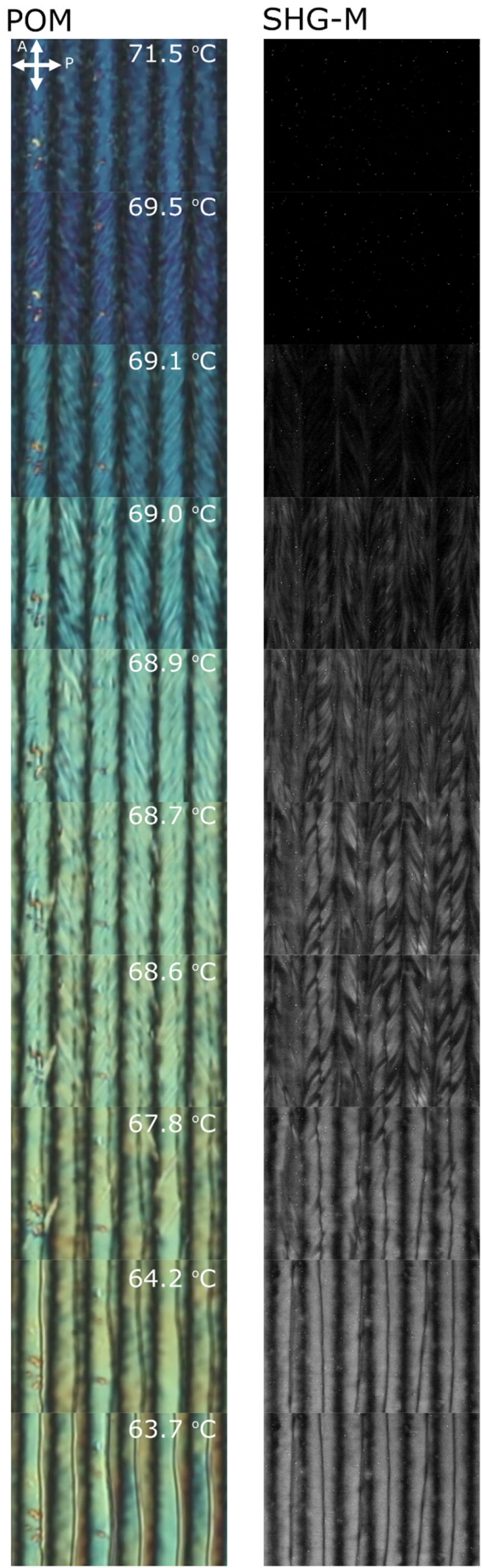

Supplementary Fig.8. Snapshots of the $N_S$-$N_F$ transition in a periodic Splay pattern with maximum angle of 40 degrees and a period P=40 μm as observed by POM (left) and SHG-M (right) in a 3 μm cell with one of the substrates with an inter-digitated ITO patterned electrode (equivalent to that shown in Fig.1.d in the main manuscript). The texture corresponds to an area without ITO in the inter-digitated electrode, showing that despite the absence of ITO the surface splay prescribed preferred director direction prevents the formation of random positioned domains as shown in Figure 1 in the main manuscript. $N_S$-$N_F$ pretransitional behaviour is evidenced by the stripe texture appearing overlaid on the prescribed pattern. Disclination lines appear smoothly through the transition along the full pattern in those areas where the splay changes sign (see for example temperatures 68.9 °C and 68.7 °C). Final structural relaxation can be observed in the last POM image of the sequence. On SHG-M the changes associated to it are neglectable.

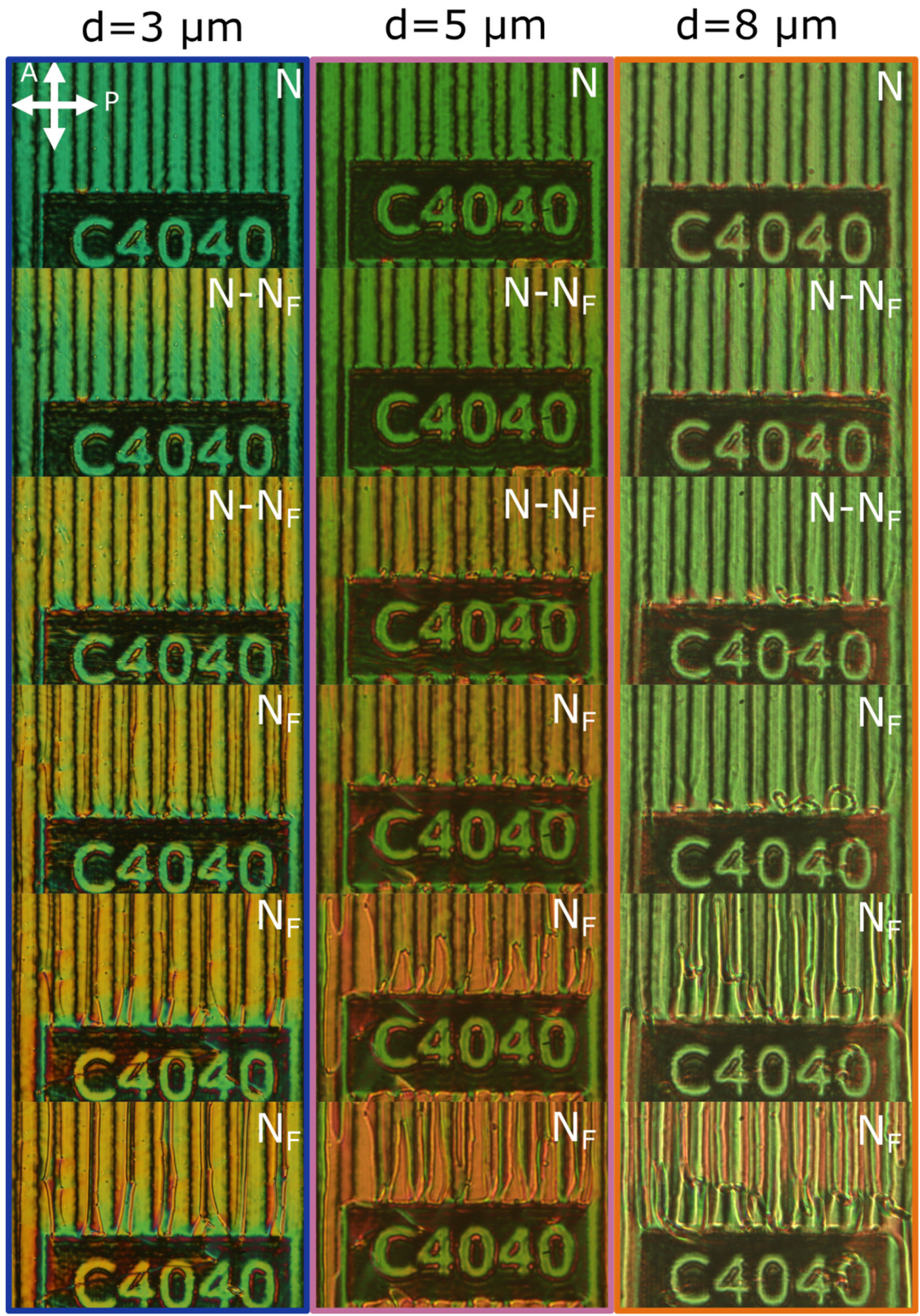

Supplementary Fig.9. Details of the N-$N_F$ transition in RM734 in the splay pattern $\vartheta = \vartheta_0 \sin(2\pi x/P)$ with $\vartheta_0 = 40°$ and $P = 40\ \mu m$ for three different cell thicknesses d= 3, 5 and 8 µm. Last two rows show the propagation of the structural relaxation. For d=3 µm, final structure resembles that of the intermediate state with the surfaces constraining the appearance of twist. In the case of the thicker cells with d= 5 and 8 µm clear differences can be observed, with the final structural relaxation involving the appearance of more complex structures.

**Polarization patterning in ferroelectric nematic liquids**

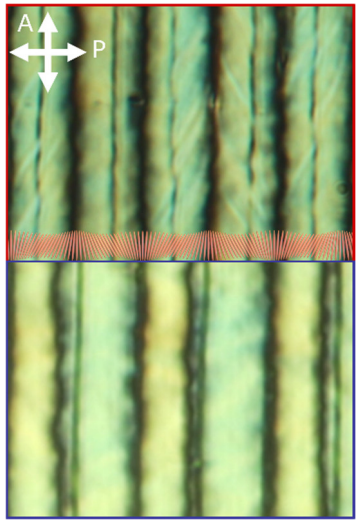

Supplementary Fig.10. Different positions of the disclination lines with respect to the photopatterned splay structure as observed in 3.06 µm cell.

## Supplementary Note V - Dtmm transmission spectra simulations

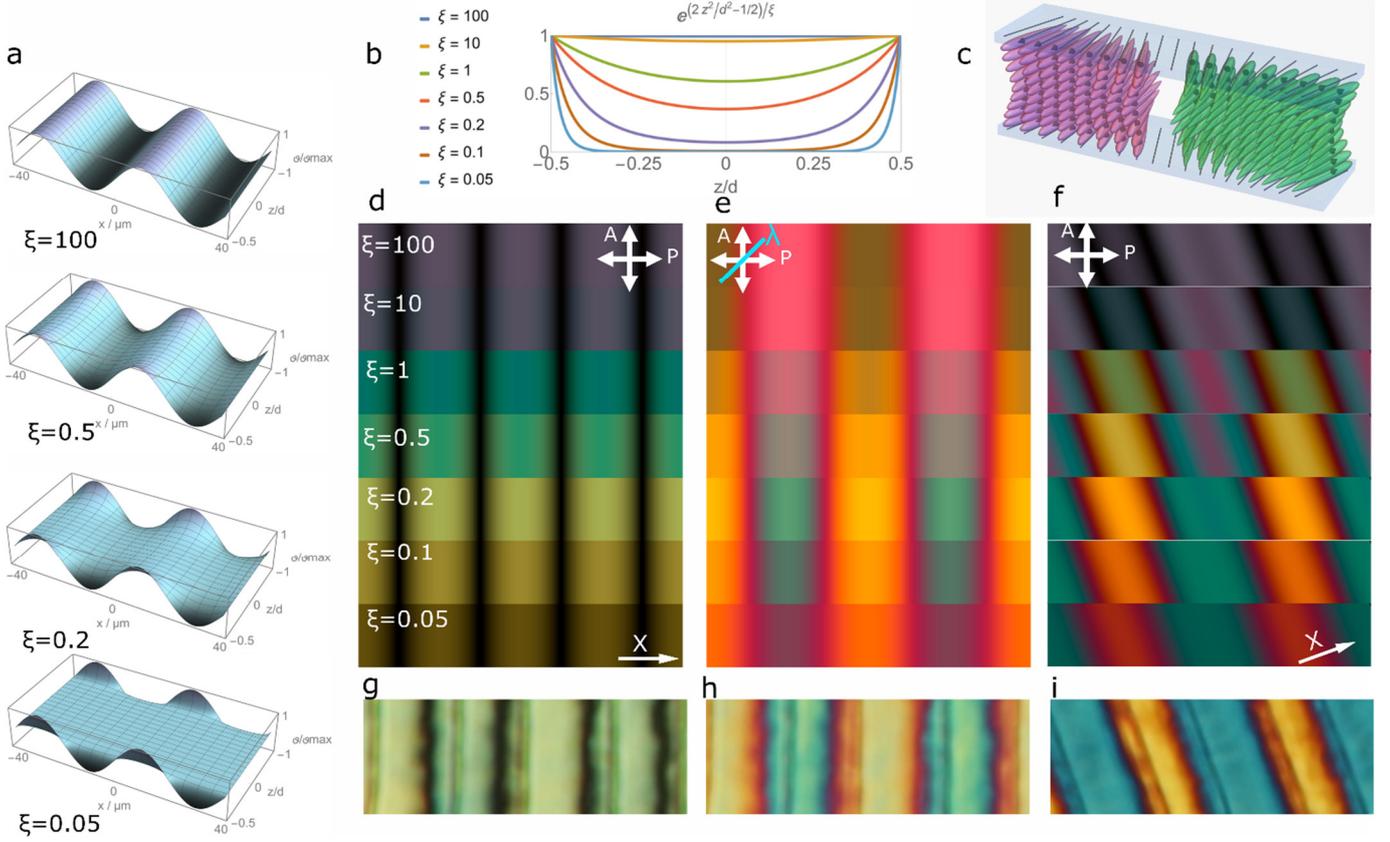

Supplementary Fig.11. Effect of twist thickness on calculated transmitted intensity patterns by dtmm considering a splay pattern $\vartheta_{surf} = \vartheta_0 \sin\left(\frac{2\pi x}{P}\right)$ with maximum angle $\vartheta_0 = 40°$, periodicity $P = 40\ \mu m$, $\Delta n = 0.19$, where $n_o$ is taken to be 1.52 and the thickness of the cell is 3.06 µm. The most clear evidence showing that the inscribed surface splay decreases in amplitude towards the centre of the cell arises from the optical transmission texture obtained when rotating the sample. While for a uniform splay structure across the cell ($\xi = 100$), no alternating colours are expected, experiments clearly show blue and yellow areas. Such behaviour, and complementary conditions, can be well simulated with dtmm considering a structure in which the director "unsplays" towards the cell centre according to $\vartheta = \vartheta_{surf} e^{(2z^2/d^2 - 1/2)/\xi}$, with $\xi = 0.2$ and being $\vartheta = 0°$ along the periodic lines. Corresponding dtmm simulations at different conditions can be found in Supplementary Fig.15. (a) Angle profile across the cell thickness and two splay periods for different $\xi$ values. (b) Profile of $e^{(2z^2/d^2 - 1/2)/\xi}$ across the cell thickness for the $\xi$ values used later in (d-f). (c) Schematic representation of the director structure inside the confining cell. (d-f) Dtmm transmission spectra simulations for different $\xi$ values for a sample with the lines aligned with respect the crossed polarizers (d), with an additional lambda plate inserted at 45° as indicated by the blue line (e) and for the sample rotated 20 degrees with respect to the crossed polarizers (f). (g-i) Experimental POM images of the periodic splay pattern obtained for DIO in the same conditions as those simulated.

**Polarization patterning in ferroelectric nematic liquids**

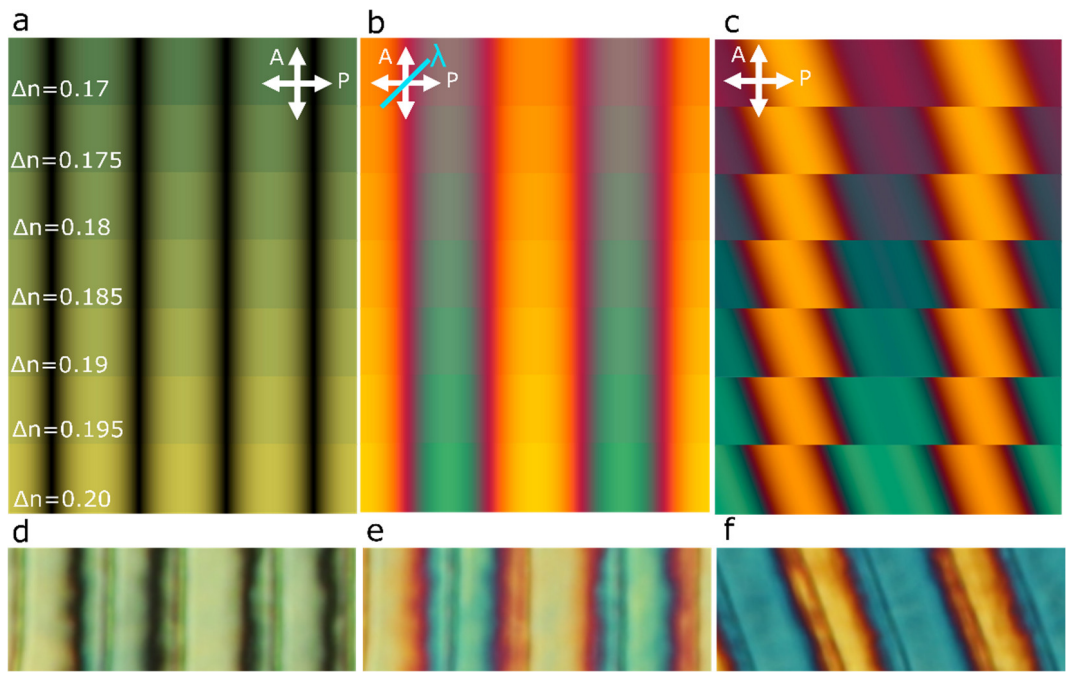

Supplementary Fig.12. Calculated transmitted intensity splay patterns for a maximum splay angle of 40° and $\xi = 0.2$ for different birefringence values. The ordinary refractive index value n$_o$ is taken to be 1.52 and the thickness of the cell is 3.06 µm. Images (a-c) show the different dtmm simulations for three different conditions: (a) pattern aligned with crossed polarizers, (b) with a lambda plate inserted at 45° as indicated by the blue line and (c) for the sample rotated 20 degrees with respect to the crossed polarizers. (d-f) Show the corresponding POM images of the periodic patterns in DIO.

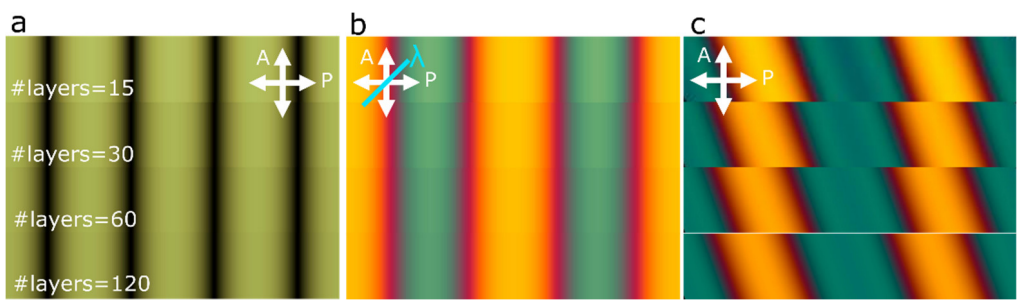

Supplementary Fig.13. Calculated transmitted intensity splay patterns for a splay maximum angle of 40°, $\xi = 0.2$ and $\Delta n = 0.19$ considering different number of discretization layers for a 3.06 µm cell.

**Polarization patterning in ferroelectric nematic liquids**

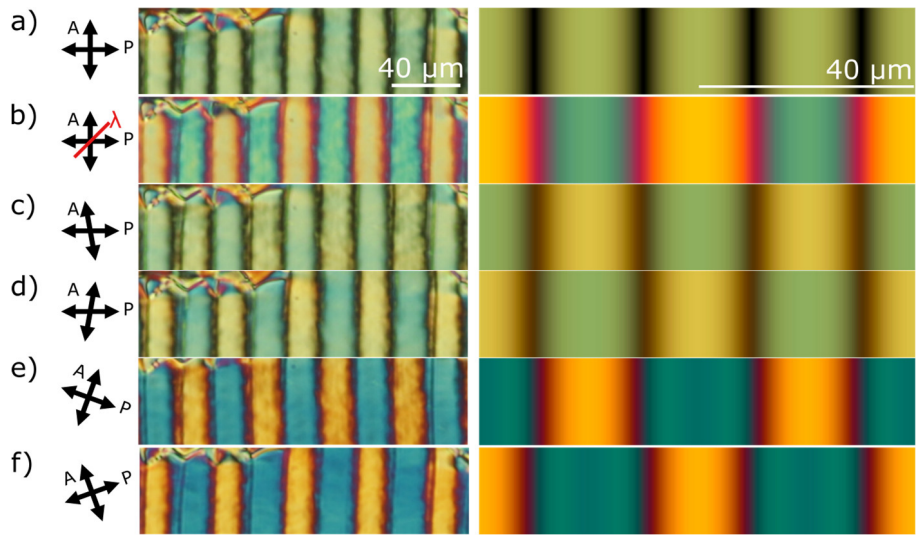

Supplementary Fig.14. DIO: Comparison of POM images and dtmm simulations for all the studied geometries for the splay pattern with $\vartheta_0 = 40°$, $\xi = 0.2$ and P= 40 $\mu m$ in a 3.06 µm cell. Dtmm simulations were performed considering $\Delta n = 0.19$, where n$_o$ is taken to be 1.52. a) Crossed polarizers along the pattern, b) with lambda plate, c&d) uncrossing analyser in opposite directions and e&f) rotating the sample in opposite directions.

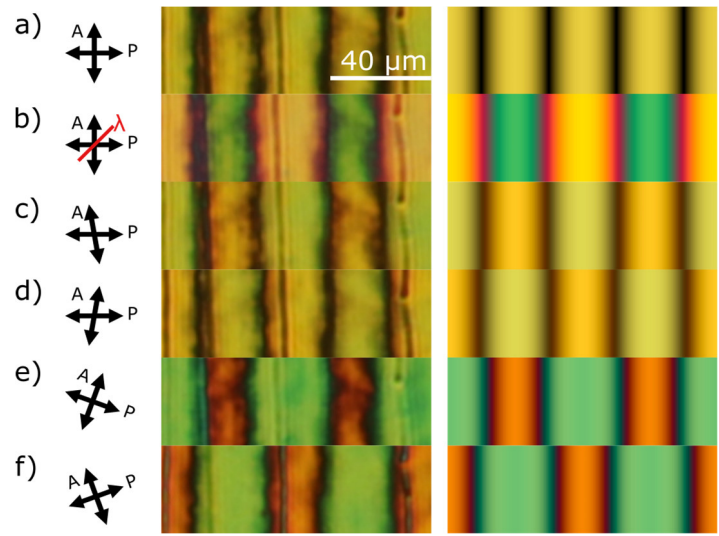

Supplementary Fig.15. RM734: Comparison of POM images and dtmm simulations for all the studied geometries for the splay pattern with $\vartheta_0 = 40°$ $\xi = 0.2$, and $P = 40\ \mu m$ in a 3.06 μm cell. Dtmm simulations were performed considering $\Delta n = 0.21$, where $n_o$ is taken to be 1.52. a) Crossed polarizers along the pattern, b) with lambda plate, c&d) uncrossing analyser in opposite directions and e&f) rotating the sample in opposite directions.

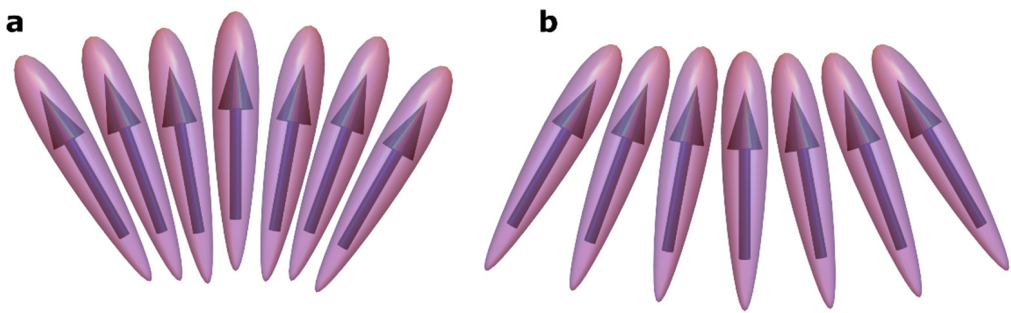

Supplementary Fig.16 For a polar nematic phase, made of asymmetric molecules, due to minimization of excluded volume, polarization and deformation are coupled. That is, there is a favourable splay (a) and a non favourable splay (b) with respect to polarization direction.

## Supplementary Note VI – Polarization guiding structures

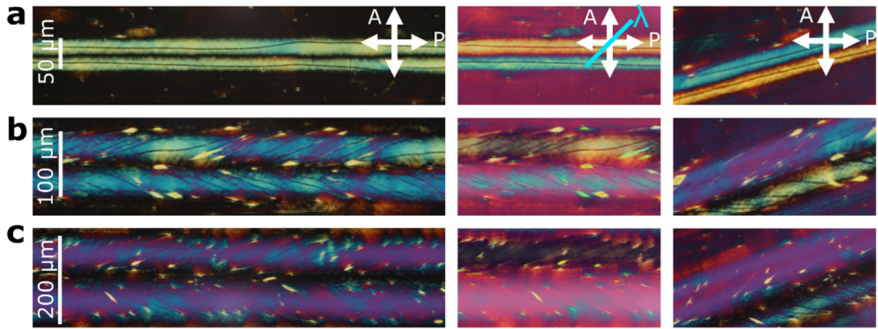
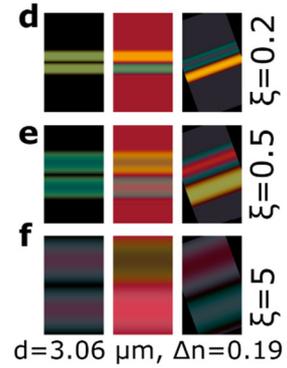
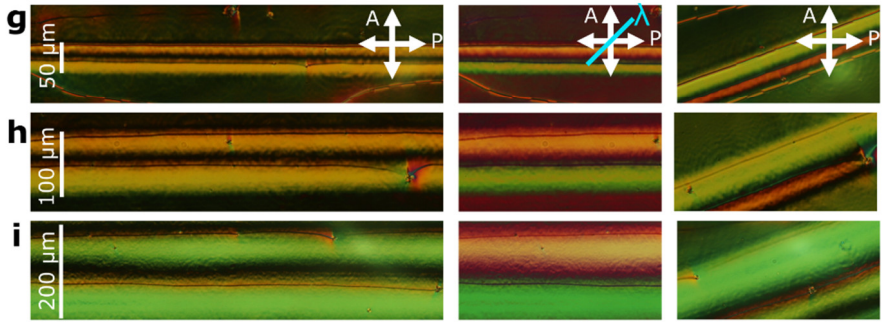
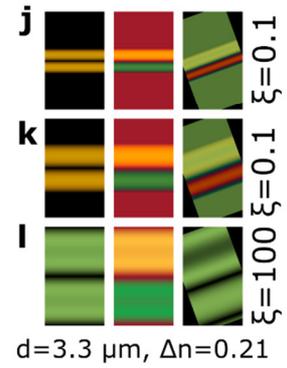
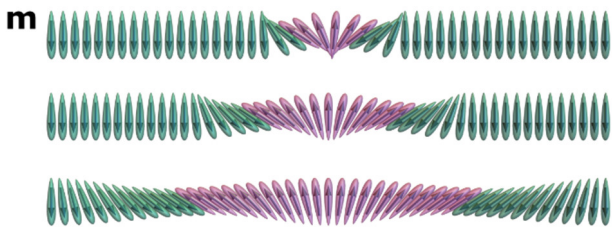
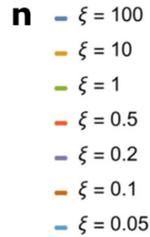
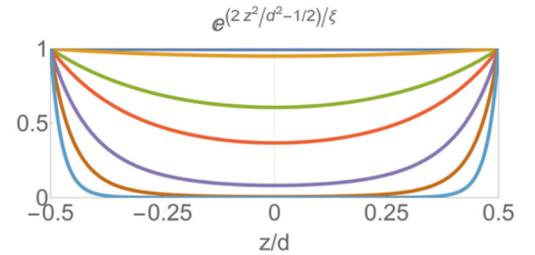

Supplementary Fig.17 Comparison of POM observations and dtmm transmission spectra simulations for the different single splay lines in DIO (a-f) and RM734 (g-l). For dtmm simulations a uniform background was considered in which single splay lines of maximum splay angle $\vartheta_0 = 45°$ and period $P$= 50, 100 and 200 μm are embedded. In the splay regions, the azimuthal angle "unsplays" as $\vartheta = \vartheta_{surf} e^{(2z^2/d^2-1/2)/\xi}$, being $\vartheta = 0°$ along the splay lines. Simulations were performed considering $\Delta n = 0.19$ for DIO and $\Delta n = 0.21$ for RM734 and the employed structural parameters for each simulation are shown in the figure. Results show that while in DIO, only the larger splay curvature ($\vartheta_0 k = 0.1$, for P=50 μm) results in a controlled polarization patterning as evidenced by the large amount of defect lines in b and c. In RM734 the three splay curvatures ($\vartheta_0 k = 0.1$, 0.05 and 0.025 μm$^{-1}$, i.e. approx. 6, 3 and 1.5 μm) result in well-oriented samples. Interestingly, it is possible to discern how the decrease of the splay curvature, implies in both cases, the reduction of depolarization field as evidenced by the decrease of the "unsplay" structure. m) Schematic representation of the three considered structures. n) Profile of the $e^{(2z^2/d^2-1/2)/\xi}$ factor diminishing the splay across the cell thickness.

**Polarization patterning in ferroelectric nematic liquids**

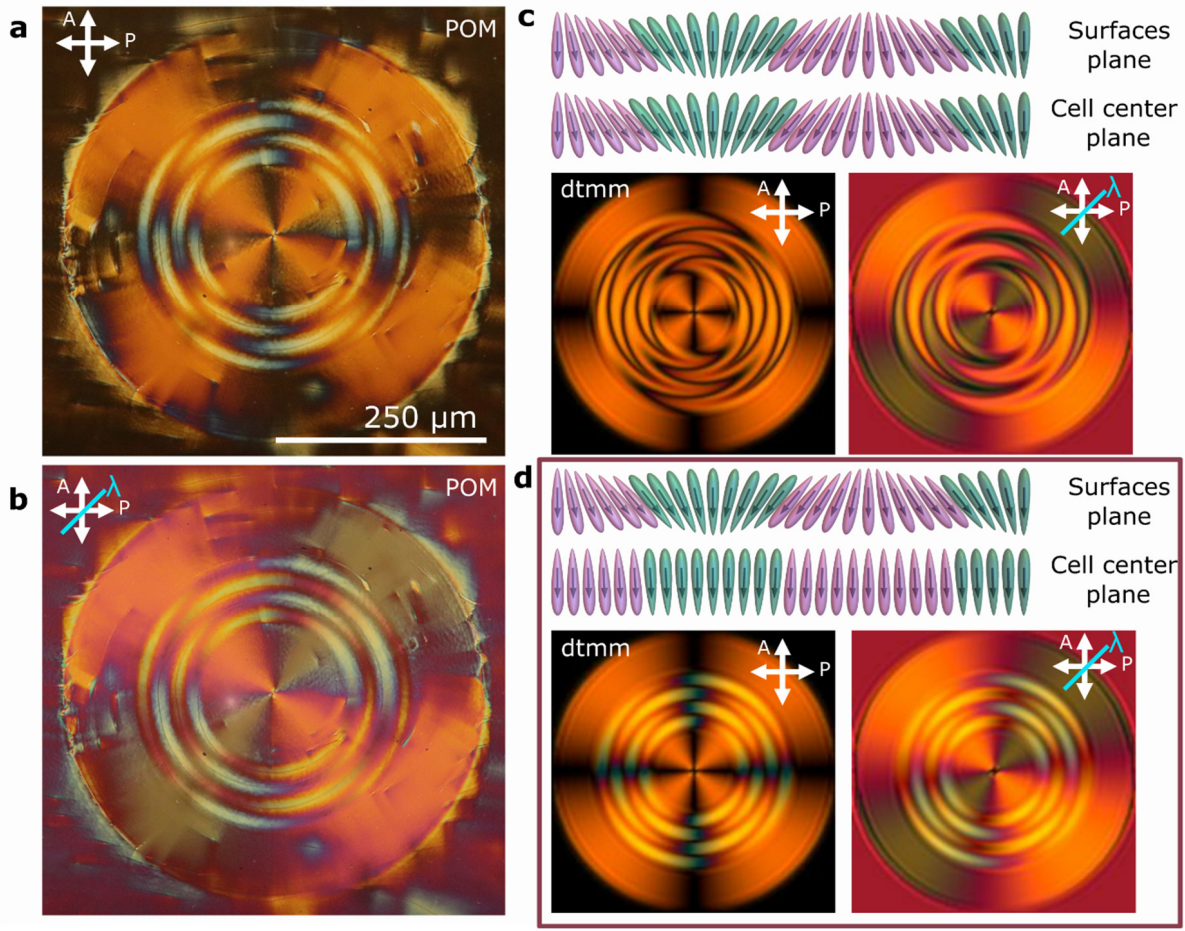

Supplementary Fig.18. Comparison of POM observations (a&b) and dtmm transmission spectra simulations (c&d) for DIO in a structured bend circle in which several splay lines are tangentially embedded. POM texture a) under crossed polarizers and b) with a full lambda plate inserted at 45 degrees as indicated by the blue line. Dtmm simulations were performed considered a uniform splay across the cell thickness d (c) and (d) a structure in which azimuthal angle varies as $\vartheta = \vartheta_{surf} e^{(2z^2/d^2 - 1/2)/\xi}$ with $\xi = 0.2$ in the splay regions and considering $\Delta n = 0.19$.

## Supplementary Note VII – Model for the structure of the splay patterns:

In an apolar nematic liquid crystal, the structure of the director field is determined by a balance of elastic, electric/magnetic and surface torques[5]. In a ferroelectric NLC, the field contribution includes also so called depolarization field, which is electric field caused by the ferroelectric body itself. To assess the stability of the structures deduced from POM in the case of periodic splay pattern, we used a simplified model.

First, the local field **E** was calculated using the following expression for the director field **n(r)** within one half of a periodic splay pattern:

$$\mathbf{n}(x,z) = (\sin\vartheta(x,z), \cos\vartheta(x,z)\cos\varphi(x,z), \cos\vartheta(x,z)\sin\varphi(x,z)) \qquad (1)$$

$$\mathbf{P}_s = P_0 \mathbf{n} \qquad (2)$$

$$\vartheta_i(x,z) = A_0 \sin\frac{k_x x}{2} e^{(2z^2-0.5d^2)/(\xi d^2)}\left(1 - e^{(2x^2-0.5L^2)/(0.05L^2)}\right), \qquad (3)$$

$$\varphi_i(x,z) = 0 \qquad (4)$$

where $d$ is the thickness for the LC cell, $\xi = 0.2$ as descried in the manuscript, $k_x = \frac{2\pi}{L}$, and $L$ a half of the splay period $P$. In the other half of the pattern $\vartheta_1(x,z) = \vartheta(x,z) + \pi$ and $\varphi_1(x,z) = \pi - \varphi(x,z)$. At the cell surface boundaries ($z = \pm d/2$), the orientation of **n** is prescribed by the photo-pattern with $\vartheta_s(x) = A_0 \sin\frac{k_x x}{2}\left(1 - e^{(2x^2-0.5L^2)/(0.05L^2)}\right)$, with the last term being discussed below. To simplify calculations, it was assumed the dielectric tensor is isotropic, $\boldsymbol{\varepsilon} = \varepsilon \mathbf{I}$.

In the case when there are no free ions, the local field is the depolarization field, which can be calculated using bound charges for being its source. The bound charge has two contributions, volume charge $\rho_b = -\nabla \cdot \mathbf{P}_s$, and surface charge $\sigma_b = \mathbf{v} \cdot \mathbf{P}_s$, where **v** is a vector normal to the surface of the ferroelectric. The Eq.3 differs from the assumed structure in dtmm simulations in the last factor, which is used to transform surface bound charge in the edges of the splay to a smooth charge distribution within a thin layer ($\approx 0.025L$) at the boundary of our calculation box, so that $\sigma_b = 0$. That is, cell surface boundaries are considered as prescribed by photopatterning everywhere expect in a thin layer ($\approx 0.025L$) around $x = \pm L/2$ where the splay changes sign. The electrostatic potential $\Phi(\mathbf{r})$ is calculated using the Poisson equation

$$\nabla^2 \Phi_{dep} = -\frac{\rho_b}{\varepsilon\varepsilon_0} \qquad (5)$$

in a box ($-P/4 \leq x \leq P/4, -d/2 \leq z \leq d/2$), which is periodic in the $x$-direction and finite in the $z$-direction. Because $\rho_b$ is the same in both halves of the splay pattern, Eq.(5) needs to be solved only for half of the pattern.

In a realistic case, there are free ions present in the material, which screen the depolarization field. Assuming the free charge density follows Boltzmann distribution, $\rho^\pm = \pm e n_0 \text{Exp}(\mp e\Phi/(k_B T))$, and positive and negative ions have the absolute charge, i.e., they carry a charge $e = \pm Z e_0$, (where $e_0$ is the elementary charge, $Z$ positive integer, and the number density $n_0$ is the same for positive and negative free ions), the electrostatic potential can be calculated using Poisson-Boltzmann equation

$$\nabla^2 \Phi_n = \beta^2 \sinh \Phi_n - \rho_{b,n} \qquad (6)$$

Here, $\Phi_n = e\Phi/k_B T$. If $\Phi_n < 1$, as an approximation, a linearized Poisson-Boltzmann equation

$$\nabla^2 \Phi_n = \beta^2 \Phi_n - \rho_{b,n} \qquad (7)$$

can be used to evaluate local electric fields. We used such a normalization that $\beta^2 = (2n_0 e)/(P_0 k_x A_0)$, $\rho_{b,n} = \rho_b/(P_0 k_x A_0)$, and the length is normalized to $\xi_b = \sqrt{(\varepsilon\varepsilon_0 k_B T)/(P_0 k_x A_0)}$. In this normalization, the coefficient $\beta = \xi_b/\lambda_D$, where $\lambda_D = \sqrt{(\varepsilon\varepsilon_0 k_B T)/(2n_0 e^2)}$ is the Debye length.

Two cases of boundary conditions were considered, (a) the material is in contact with grounded electrodes:

$$\Phi\left(x, \pm\frac{d}{2}\right) = 0, \tag{8}$$

or (b) with glass with the dielectric constant $\varepsilon_g$:

$$\frac{\partial \Phi}{\partial x}\left(x, \pm\frac{d}{2}\right) = \frac{\partial \Phi_{glass}}{\partial x}\left(x, \pm\frac{d}{2}\right) \tag{9}$$

$$\varepsilon\frac{\partial \Phi}{\partial z}\left(x, \pm\frac{d}{2}\right) = \varepsilon_g \frac{\partial \Phi_{glass}}{\partial z}\left(x, \pm\frac{d}{2}\right), \tag{10}$$

where $\Phi_{glass}$ is the electrostatic potential in the glass. $\rho_b$ was approximated with a finite Fourier series, and Eq.7. was solved numerically in Fourier space. The field was calculated as $\mathbf{E} = -\nabla\Phi$.

In Supplementary Fig.19 and Fig.20, the comparison of the local field is shown for the two boundary conditions and different ion concentrations for the initial structure and $\varepsilon$ =100 or 1000 respectively. In the calculations, the following values were used: $P_0 = 0.05 \frac{\text{As}}{\text{m}^2}$, $d = 3$ µm, $L = 20$ µm, $A_0 = 40°$, $\varepsilon$ =100 or 1000, $\varepsilon_0 = 8.85 \cdot 10^{-12} \frac{\text{As}}{\text{Vm}}$, $k_B T = 4 \cdot 10^{-21}$ J, $Z$=1, $e_0 = 1.6 \cdot 10^{-19}$ As and $\beta^2$ =0, 1 and 10, which correspond to ion concentrations of $n_0 = 0$ m$^{-3}$, 1.7·10$^{22}$ m$^{-3}$ and 1.7·10$^{23}$ m$^{-3}$. The comparison of the order of magnitude of the electrostatic energy density 1/2 $\mathbf{P} \cdot \mathbf{E}$ with twist elastic energy $K_2/(0.1d)^2 \sim 20$ J/m$^3$ shows that in all cases, the local fields are too large for the structure to be stable. (Here, $K_2$ is the twist elastic constant.) This means that the electric field torque acting on $\mathbf{P}$ is much larger than the nematic elastic torque, and, consequently, the structure will relax towards stable structure determined by the minimum of the free energy. To assess the difference between the initial and relaxed structure, a simplified model for free energy was used.

In general, FNLC can be described by two coupled order parameters, a nematic quadrupolar, i.e., tensor $\mathbf{Q} = S(\mathbf{n} \otimes \mathbf{n})$, and electric polarization vector $\mathbf{P}$[6]. Here, $S$ is the scalar order parameter and $\mathbf{n}$ the director (with the symmetry $\mathbf{n} \equiv -\mathbf{n}$), which denotes direction of the average orientation of the molecules in the nematic phase[5]. In the model, we made the following assumptions: (i) $S$ is constant, so the nematic order can be described only by $\mathbf{n}$; (ii) $\mathbf{P} = \mathbf{P}_s + \varepsilon_0(\boldsymbol{\varepsilon} - \mathbf{I})\mathbf{E}$, where $\mathbf{P}_s = P_0 \mathbf{n}$; (iii) $P_0$ is constant; (iv) dielectric tensor is isotropic, $\boldsymbol{\varepsilon} = \varepsilon \mathbf{I}$; and (v) the orientation of the director at the surface is the same as prescribed by photo-patterning, that is so-called strong anchoring boundary condition. Here, it has to be noted, that this simplified model is not suitable for the description of topological defects, around which neither $S$ nor $P_0$ is expected to be constant. The model is only used to assess whether this simple approach can explain the observed structures away from the defects. Additionally, while the effective value of dielectric constant measured by dielectric spectroscopy is large, i.e. of the order of 10000 (which mainly comes from the reorientation of $\mathbf{P}_s$)[6], the value of $\varepsilon$ as defined above only accounts for induced polarization and is expected to be of the order 100 – 1000. The value is larger close to the ferroelectric phase transition and it decreases with the temperature, i.e., when the system is deeper in the ferroelectric nematic phase.

Stable structures can be found by minimization of the Landau-de Gennes type of free energy functional. In the general case when free ions are present, the functional must include also the electrostatic potential[7]. However, if we assume that the dynamics of free charges is much faster than the dynamics of $\mathbf{n}$, then during the relaxation of $\mathbf{n}$, the electrostatic potential is given by the solution of the Poisson-Boltzmann equation. In such case, the relaxation method can be used to minimize the relevant part of the Landau-de Gennes type of free energy functional:

$$F_\mathbf{n} = \int \left( \tfrac{1}{2} K_1 |\mathbf{S} - \mathbf{S}_0|^2 + \tfrac{1}{2} K_2 Tw^2 + \tfrac{1}{2} K_3 |\mathbf{B}|^2 - \tfrac{1}{2} P_0 \mathbf{n} \cdot \mathbf{E} \right) dV. \tag{11}$$

The relaxation steps were performed with respect to $\vartheta(x,z)$ and $\varphi(x,z)$. At each step, $\mathbf{E}$ was recalculated using Eq.7. Here, $K_i$ ($i$ = 1,2,3) splay, twist, and bend elastic constants with corresponding deformations $\mathbf{S} = \nabla \cdot \mathbf{n}$, $Tw = \mathbf{n} \cdot (\nabla \times \mathbf{n})$, $\mathbf{B} = \mathbf{n} \times (\nabla \times \mathbf{n})$, and $\mathbf{E} = -\nabla \Phi$. The flexoelectric term is included in the first term, where $S_0 = \gamma \mathbf{P} \cdot \mathbf{n}/K_1$ is the ideal splay curvature, which would minimize the splay elastic energy. The sign of $\mathbf{S_0}$ determines the preferred direction of $\mathbf{P}$ when splay deformation is present in the system. The ideal splay curvature which would minimize the splay elastic energy, is $\mathbf{n} \cdot \mathbf{S_0}$. However, because of the assumptions (ii) and (iii) described above, $S_0$ does not enter the local relaxation equations for $\vartheta(x,z)$ and $\varphi(x,z)$. In the calculations the following values of elastic constants were used: $K_1 = K_3 = 20$ pN, $K_2 = 2$ pN.

Supplementary Fig.21 shows the local field in the case of two examples of relaxed structures. At the edges of the pattern, where defects are observed in the experiment, the local fields and deformations are large, counteracting each other. As already discussed, in this region, our simplified model is not expected to yield realistic results and we will not discuss it further. If the density of ions is sufficient ($\beta^2 \gtrsim 1$), the free ions screen the part of the depolarization field originating from the polarization in the part of the sample which is more than a few screening lengths away from a given point. Therefore, the structure at the edge has little or no effect on the structure in the middle of the pattern. In Supplementary Fig.22, the local fields away from the edges in the structure after minimization are compared with the initial one for the case screening length is 20 nm ($\beta^2 = 10, \varepsilon = 100$) and in Supplementary Fig.23 the corresponding $\vartheta(x,z)$ and $\phi(x,z)$ values of the structure across the region are shown. Although the structure differs very little from the initial one (Supplementary Fig.23), the decrease of the depolarization field is substantial, i.e., more than an order of magnitude (Supplementary Fig.22). Most notable is the appearance of out-of-plane splay deformation, which causes such a redistribution of bound and free charges (Supplementary Fig.24) that the field is reduced. It has to be noted here that the region as a whole is neutral, so the sum of bound charges and the sum of free charges are both zero. This means that independently of what the structure around the defects at the edge of the splay is, the total bound charge there is the same. Its value is equal to minus the sum of the charge in the middle of the splay structure, so the total bound charge is zero.

As shown by SHG-I, disclination lines separate regions with opposite polarization direction. As shown in Supplementary Fig.25, although appearing at the edge of the splay patterns, in some occasions they have also been observed towards the center of the splay region. This could be attributed to slight asymmetries of the photopatterned surface splay structure, which would provoke asymmetries in the otherwise evenly distributed charge density expected for perfect splay structure shown in Supplementary Fig.25. Such structure asymmetry would then cause a shift of the polarization reversal region towards the center of the splay, which carries opposite electric charge.

Due to the polarity of the phase, topological charge of the disclination lines should be an integer. Additionally, in order to avoid electrical charge in the core, twist deformations are the most favourable, and thus, a possible structure of such lines could be a twist-like disclination with topological charge ±1 as shown in Supplementary Fig.25.b. When embedded in a non-splayed uniform background, the overall structure is electrically neutral. However, when embedded in a splayed background, it carries a charge as discussed above.

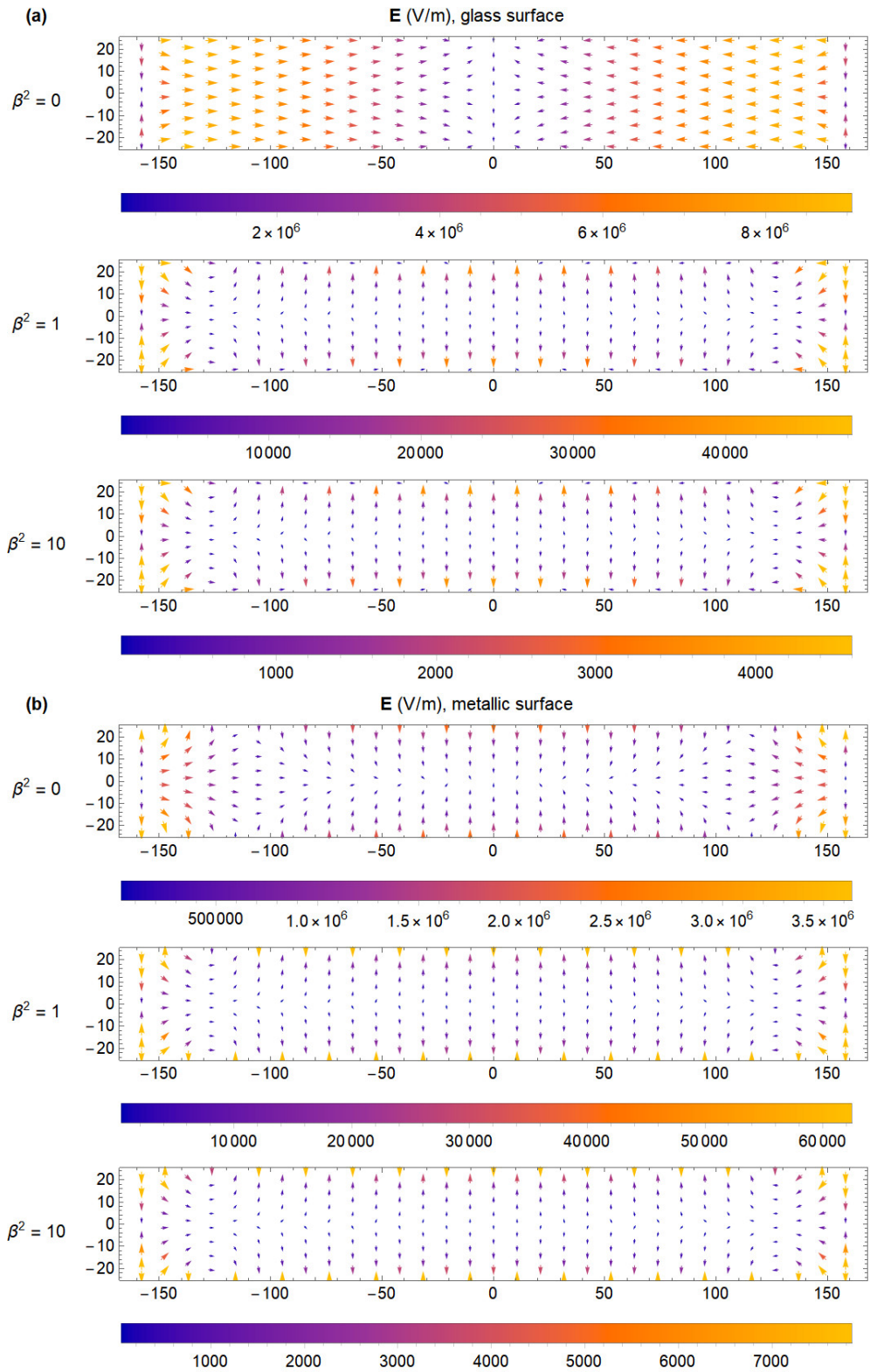

Supplementary Fig.19: xz- cross section of the local field E in units of V/m calculated for the structure given by Eqs. (1) –(3) for (a) glass surface and (b) grounded metal surfaces for $\varepsilon$ = 100, and $\beta$ as marked. The x and z coordinates are is given in units of $\xi_b = 63$ nm.

**Polarization patterning in ferroelectric nematic liquids**

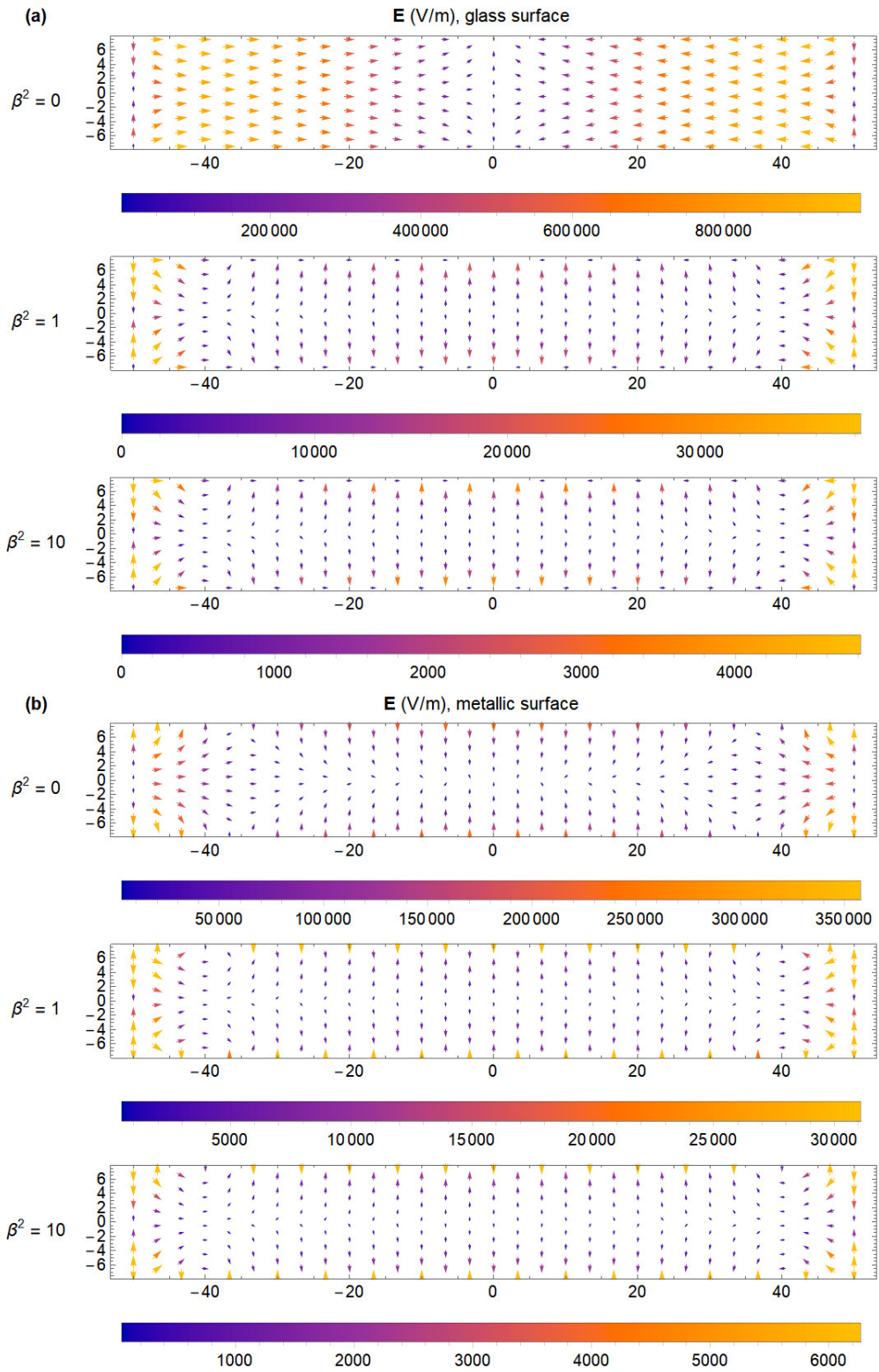

Supplementary Fig.20: xz- cross section of the local field E in units of V/m calculated for the structure given by Eqs. (1) – (3) for (a) glass surface and (b) grounded metal surfaces for $\varepsilon$ = 1000, and $\beta$ as marked. The x and z coordinates are given in units of $\xi_b = 200$ nm.

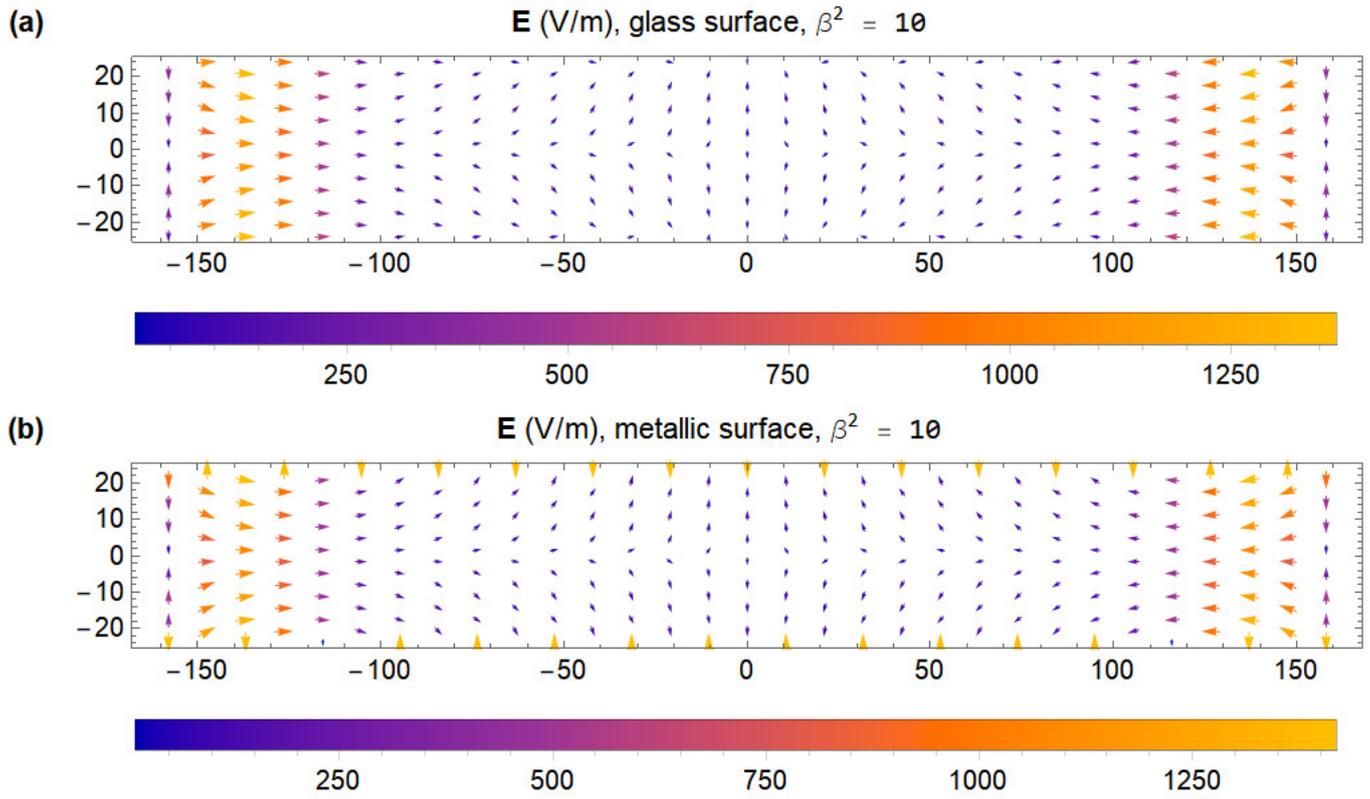

Supplementary Fig.21: xz- cross section of the local field for the structure after relaxation for $\beta^2 = 10$ and $\varepsilon = 100$. The x and z coordinates are is given in units of $\xi_b = 63$ nm.

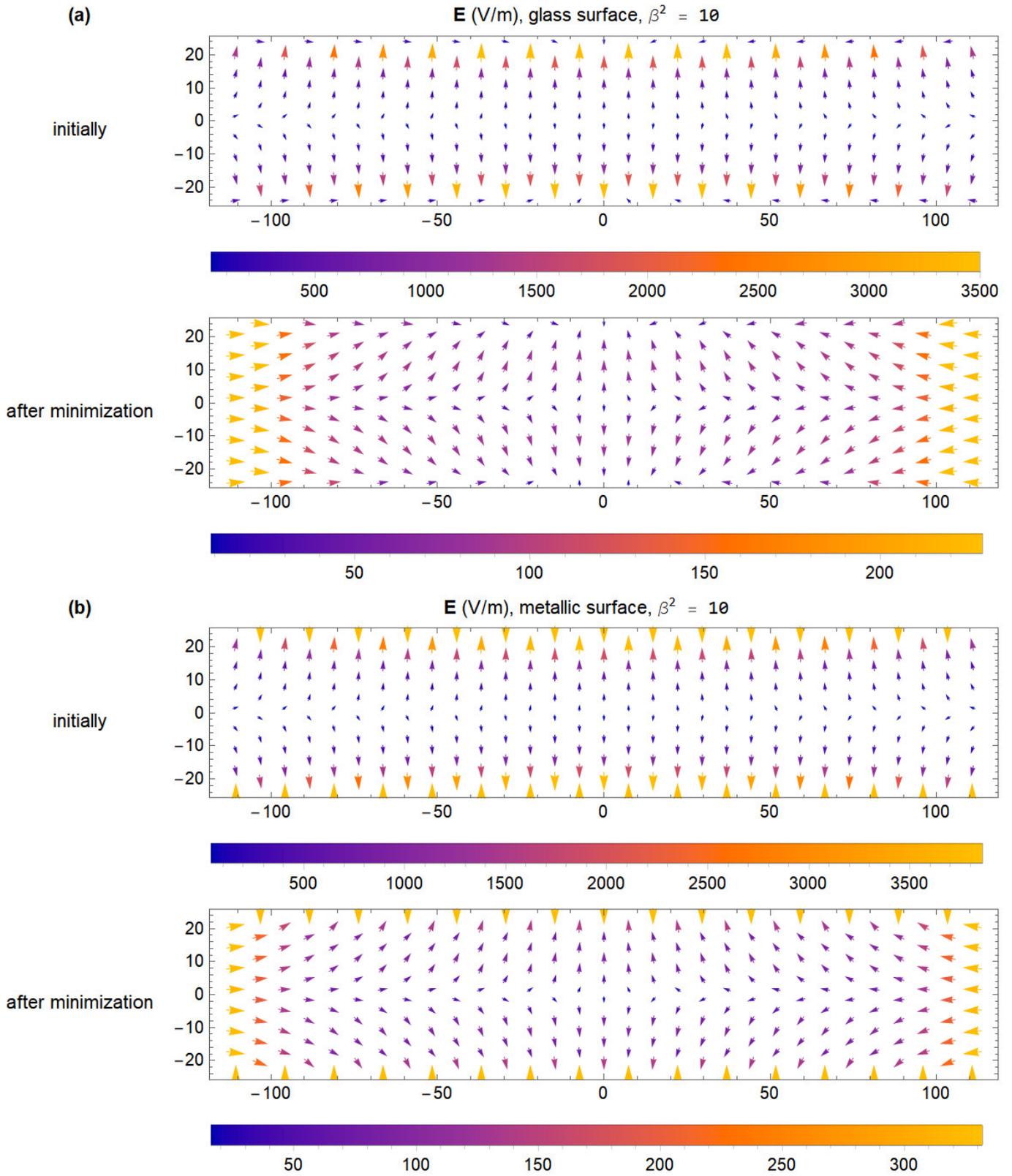

Supplementary Fig.22: Comparison of the local field for initial structure and the structure after relaxation in the central region ($-0.7\ L/2 < x < 0.7\ L/2$) for $\beta^2 = 10$ and $\varepsilon = 100$. The x and z coordinates are is given in units of $\xi_b = 63$ nm.

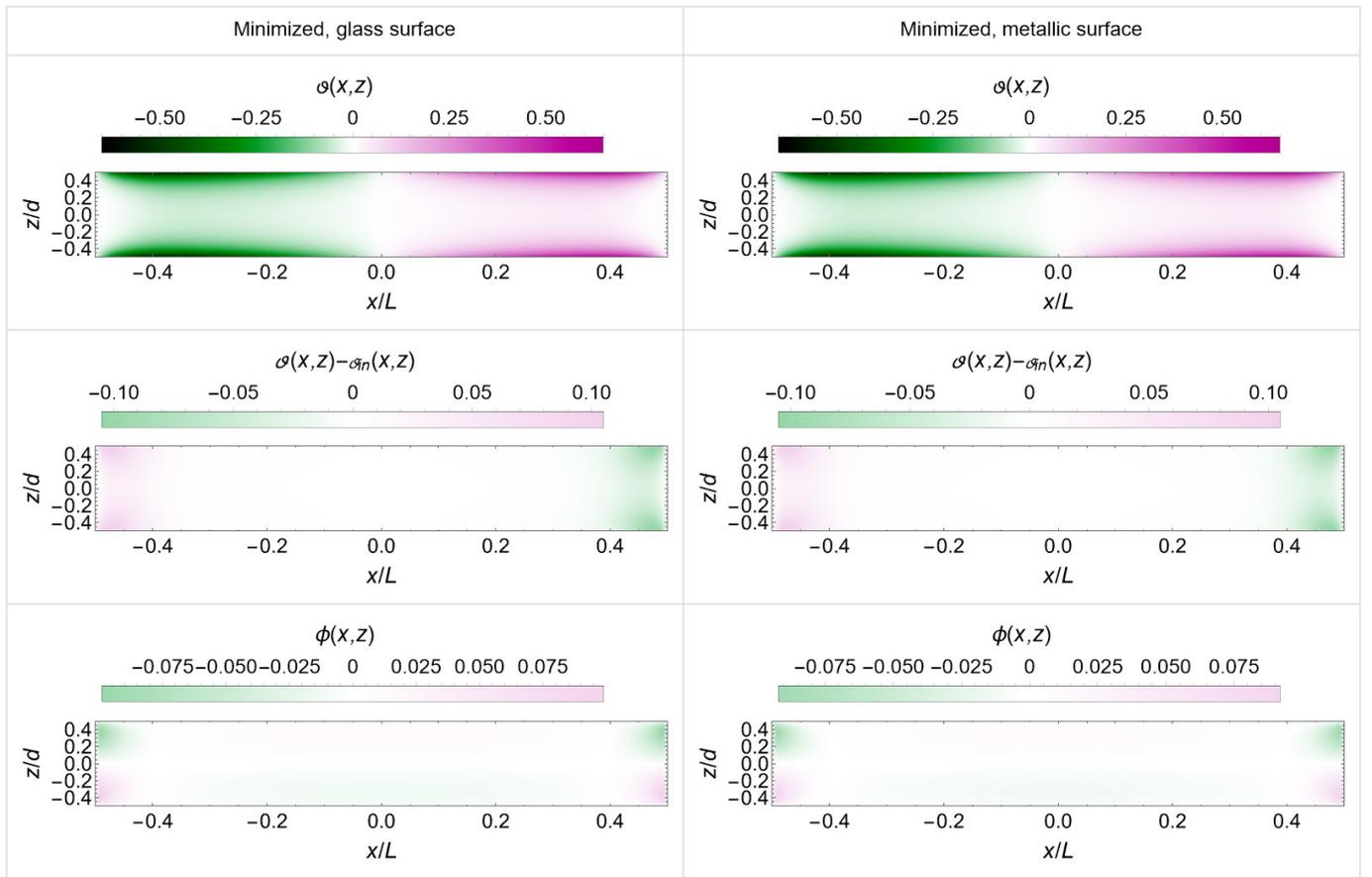

Supplementary Fig.23: Comparison of the structure before and after relaxation in the central region for both types of the surface for $\beta^2 = 10$ and $\varepsilon = 100$. The initial structure is given by angles $\vartheta_{in}(x,z)$ and $\phi_{in}(x,z) = 0$ as described in Eq.3 and Eq.4.

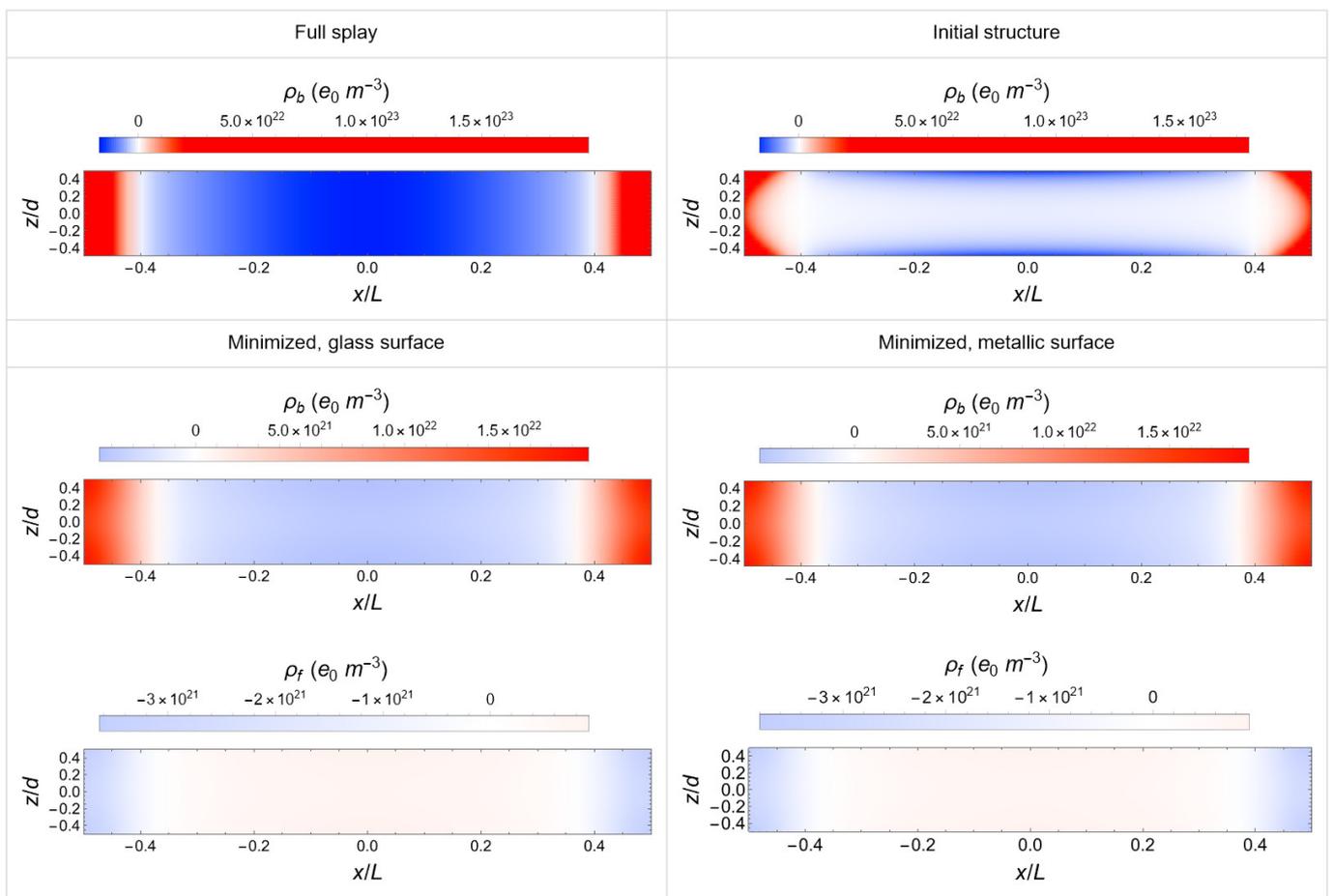

Supplementary Fig.24: Comparison of bound charge density $\rho_b$ for fully splayed structure (i.e. splay deformation independent of z), the initial unsplayed structure and relaxed structures for both boundary conditions. For relaxed structures also free charge density $\rho_f$ is shown.

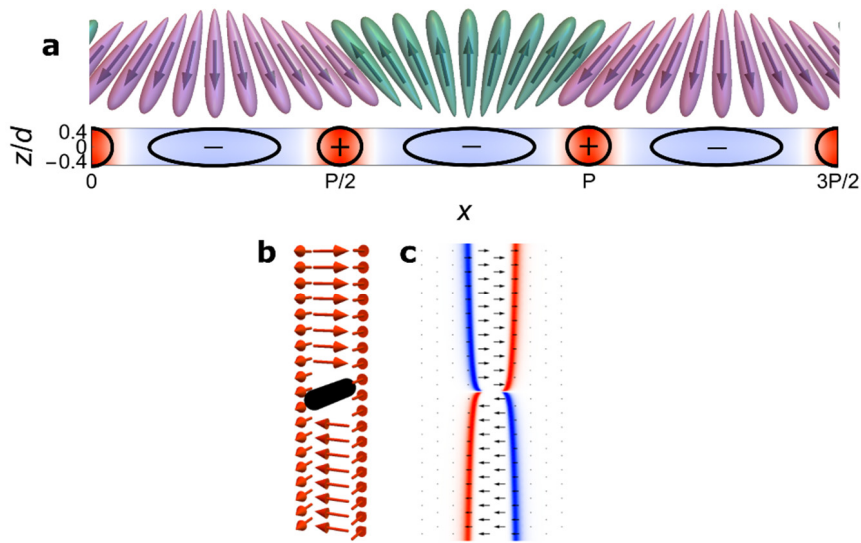

Supplementary Fig.25. Twist-like disclination lines representation. Schematic of proposed twist polarization deformation around disclination lines is shown in (b). Front view, together with the calculated bound charge density highlighted in red and blue is depicted in (a). Similarly, the bound charge density in the two planes marked in (b) can be observed in (c).

**Polarization patterning in ferroelectric nematic liquids**# References

1. Li, J. *et al.* Development of ferroelectric nematic fluids with giant-ε dielectricity and nonlinear optical properties. *Science Advances* **7**, eabf5047 (2021).
2. Mandle, R. J., Cowling, S. J. & Goodby, J. W. Rational Design of Rod-Like Liquid Crystals Exhibiting Two Nematic Phases. *Chem. Eur. J.* **23**, 14554–14562 (2017).
3. Nys, I., Berteloot, B., Beeckman, J. & Neyts, K. Nematic Liquid Crystal Disclination Lines Driven by A Photoaligned Defect Grid. *Advanced Optical Materials* **10**, 2101626 (2022).
4. Lovšin, M. & et al. *in preparation* (2023).
5. de Gennes, P. G. & Prost, J. *The Physics of Liquid Crystals*. (Clarendon Press, 1995).
6. Sebastián, N. *et al.* Ferroelectric-Ferroelastic Phase Transition in a Nematic Liquid Crystal. *Phys. Rev. Lett.* **124**, 037801 (2020).
7. Everts, J. C. & Ravnik, M. Ionically Charged Topological Defects in Nematic Fluids. *Phys. Rev. X* **11**, 011054 (2021).
36